\documentclass[%
 reprint,
superscriptaddress,
amsmath,amssymb,
aps,
]{revtex4-2}

\usepackage{graphicx}
\usepackage{dcolumn}
\usepackage{bm}
\usepackage{microtype}
\usepackage{url}
\usepackage{xcolor}
\usepackage{color}
\definecolor{amaranth}{rgb}{0.9, 0.17, 0.31}
\definecolor{palatinateblue}{rgb}{0.15, 0.23, 0.89}
\definecolor{radicalred}{rgb}{1.0, 0.21, 0.37}
\usepackage{hyperref}
\hypersetup{ linktoc=all,
	colorlinks, linkcolor={palatinateblue},
	citecolor={radicalred}, urlcolor={amaranth}
}

\usepackage{mathrsfs}

\usepackage{cleveref}

\begin{document}
\preprint{APS/123-QED}
\title{Constrain the linear scalar perturbation theory of Cotton gravity}

\author{Pengbo Xia}
\email{xpbxpb123@mail.ustc.edu.cn}
\affiliation{Deep Space Exploration Laboratory/School of Physical Sciences, University of Science and Technology of China, Hefei, Anhui 230026, China}
\affiliation{CAS Key Laboratory for Researches in Galaxies and Cosmology/Department of Astronomy, School of Astronomy and Space Science, University of Science and Technology of China, Hefei, Anhui 230026, China}

\author{Dongdong Zhang}
\email{don@mail.ustc.edu.cn}
\affiliation{Deep Space Exploration Laboratory/School of Physical Sciences, University of Science and Technology of China, Hefei, Anhui 230026, China}
\affiliation{CAS Key Laboratory for Researches in Galaxies and Cosmology/Department of Astronomy, School of Astronomy and Space Science, University of Science and Technology of China, Hefei, Anhui 230026, China}
\affiliation{Kavli IPMU (WPI), UTIAS, The University of Tokyo, Kashiwa, Chiba 277-8583, Japan}

\author{Xin Ren}
\email{rx76@ustc.edu.cn}
\affiliation{Deep Space Exploration Laboratory/School of Physical Sciences, University of Science and Technology of China, Hefei, Anhui 230026, China}
\affiliation{CAS Key Laboratory for Researches in Galaxies and Cosmology/Department of Astronomy, School of Astronomy and Space Science, University of Science and Technology of China, Hefei, Anhui 230026, China}
\affiliation{Department of Physics, Tokyo Institute of Technology, Tokyo 152-8551, Japan}

\author{Bo Wang}
\email{ymwangbo@ustc.edu.cn}
\affiliation{Deep Space Exploration Laboratory/School of Physical Sciences, University of Science and Technology of China, Hefei, Anhui 230026, China}
\affiliation{CAS Key Laboratory for Researches in Galaxies and Cosmology/Department of Astronomy, School of Astronomy and Space Science, University of Science and Technology of China, Hefei, Anhui 230026, China}

\author{Yen Chin \surname{Ong}}
\email{ycong@yzu.edu.cn}
\affiliation{Center for Gravitation and Cosmology, College of Physical Science and Technology, Yangzhou University, \\180 Siwangting Road, Yangzhou City, Jiangsu Province  225002, China}
\affiliation{Shanghai Frontier Science Center for Gravitational Wave Detection, School of Aeronautics and Astronautics, Shanghai Jiao Tong University, Shanghai 200240, China}

\date{\today}

\begin{abstract} 
We perform a cosmological test of Cotton gravity, which describes gravity by cotton tensor. We assume in Cotton gravity the background evolution is the same as the flat FLRW background. We derive the cosmological perturbation theory of the scalar mode at the linear level, where the difference from the $\Lambda$CDM model is characterized by the parameter $\beta$. We incorporate Cotton gravity with a neutrino model and perform a Monte Carlo Markov Chain (MCMC) analysis using data from the Cosmic Microwave Background (CMB) and Sloan Digital Sky Survey (SDSS). The analysis constrains parameter $\beta=-0.00008^{+0.00080}_{-0.00104}$ at the 1-$\sigma$ confidence level. We conclude that currently, there is no obvious deviation between Cotton gravity and the $\Lambda$CDM model in the linear cosmological perturbation level for observations.
\end{abstract}

\maketitle


\section{\label{sec:INTRODUCTION}INTRODUCTION}
Einstein proposed general relativity (GR) in 1915 to describe the relationship between matter and spacetime. Since then, GR has been tested at various scales \cite{Will:2014kxa,Ishak:2018his,Berti:2015itd} and has successfully explained numerous phenomena. As astronomy progresses, a multitude of observations at large scale have been made. Observations such as CMB \cite{WMAP:2003zzr,Planck:2018vyg} and galaxy rotation curve \cite{Rubin:1980zd,Lelli:2016zqa} have prompted the proposal of dark matter. Additionally, data from Super-Nova Ia \cite{SupernovaSearchTeam:1998fmf,Riess:2021jrx}, along with CMB and Baryon Acoustic Oscillations (BAO) \cite{DESI:2024mwx,SDSS:2009ocz}, provide evidence that a cosmological constant needs to be introduced to account for the universe's expanding history. With these observations, the standard cosmological model, known as the $\Lambda$CDM model, is established, describing the universe's evolution from inflation to the present era. However, many unresolved issues remain, such as understanding the physics behind dark matter and dark energy, and the recent tensions within the $\Lambda$CDM model \cite{DiValentino:2021izs}. Consequently, various modified gravity theories have been proposed to address these issues, including scalar-tensor field theory \cite{Horndeski:1974wa,Khoury:2003rn,Langlois:2015cwa,BenAchour:2016cay}, $f(T)$ gravity \cite{Cai:2015emx,Krssak:2018ywd,Cai:2011tc,Bengochea:2008gz}, and $f(R)$ gravity \cite{Sotiriou:2008rp,Nojiri:2010wj,Nojiri:2005vv,Yang:2024kdo}.

In 2021, Harada proposed a new modified gravity theory, known as ``Cotton gravity", which describes gravity using the cotton tensor, named after Émile Cotton \cite{cotton1899varietes}, instead of the Einstein tensor \cite{Harada:2021bte}. Harada demonstrated that Cotton gravity encompasses all the solutions of GR and additionally exhibits non-trivial solutions distinct from GR. Subsequently, another form of the field equation in Cotton gravity was introduced in a paper by Mantica et al. \cite{Mantica:2022flg}, which allows a clearer comparison with GR.
Harada's paper \cite{Harada:2021bte} introduced the first Schwarzschild-like metric to solve the vacuum spherically symmetric case within the framework of Cotton gravity. This solution has been generalized in a subsequent paper by Gogberashvili et al. \cite{Gogberashvili:2023wed}, suggesting significant deviations from GR. Furthermore, the effect of Cotton gravity in the galactic scale can explain the rotation curve without dark matter \cite{Harada:2022edl}. One of the motivations of our paper is to investigate whether Cotton gravity can serve as a substitute for dark matter on large scales.
 
Recently, the cosmic background evolution in Cotton gravity has been investigated in these papers \cite{Sussman:2023wiw,Sussman:2023eep,Mantica:2023ssd}. However, there is still an ongoing debate about the theoretical structures of Cotton gravity \cite{Clement:2023tyx,Sussman:2024iwk,Clement:2024pjl,Sussman:2024qsg}, which raised doubts regarding whether Cotton gravity is predictive. In our work, we will simply set the background evolution in Cotton gravity to be the same as that of the concordance $\Lambda$CDM cosmology with zero spatial curvature. (the philosophy is that since $\Lambda$CDM is well-tested, any deviation must already be small at the background level), and examine its influence at the perturbation level. We emphasize that this is a phenomenological approach.

As the observation instrument advances, we can now constrain the cosmological parameters with unprecedented accuracy. This allows for a deeper understanding of perturbations during inflation and their evolution throughout the universe's history.  Also it provides a window for studying physics like neutrinos, dark matter and dark energy \cite{Zhang:2021ecp,Brust:2013ova,Marsh:2015xka,Komatsu:2022nvu}. In our paper, we utilize the precise Planck \cite{Planck:2018vyg} and SDSS \cite{SDSS:2008tqn} data to constrain Cotton gravity and seek any signature beyond GR. To achieve this, we calculate the scalar perturbation theory of Cotton gravity and clarify its influence. Furthermore, we modify the Boltzmann code and derive the constraints on the model parameters in Cotton gravity. 

The paper is organised as follows. In Section \ref{sec:Cotton gravity}, we first introduce Cotton gravity and present the two equivalent forms of Cotton gravity field equation. In Section \ref{cosmological background and perturbations}, we discuss the evolution of cosmological background and derive the scalar perturbation evolution. We also analyze its influence. In Section \ref{CMB constraints on the parameters}, we illustrate said influence graphically and obtain the constraints on the parameters. We find that neutrinos will significantly impact the constraint results. We conclude in Section \ref{conclusion}.

\section{Cotton gravity\label{sec:Cotton gravity}}
Cotton gravity employs the cotton tensor to describe the gravitational field instead of Einstein tensor. In Cotton gravity, the field equation is given by \cite{Harada:2021bte}
 \begin{equation}
 \label{eq:cottonfieldequation}
     C_{\nu\rho\sigma}=16\pi G\nabla_{\mu}T^{\mu}{}_{\nu\rho\sigma},
 \end{equation}
 where 
 \begin{equation}
      {C}_{\nu\rho\sigma}=\nabla_{\rho}{R}_{\nu\sigma}-\nabla_{\sigma}R_{\nu\rho}-\frac{1}{6}(g_{\nu\sigma}\nabla_{\rho}R-g_{\nu\rho}\nabla_{\sigma}R), 
 \end{equation}
and
  \begin{equation}
     \nabla_{\mu}T^{\mu}{}_{\nu\rho\sigma}=\frac{1}{2}(\nabla_{\rho}T_{\nu\sigma}-\nabla_{\sigma}T_{\nu\rho})-\frac{1}{6}(g_{\nu\sigma}\nabla_{\rho}T-g_{\nu\rho}\nabla_{\sigma}T),
 \end{equation}
in which $T_{\mu\nu}$ is the usual energy-momentum tensor, and $T=T^\mu_{~\mu}$ its contraction.
In Cotton gravity, the conservation law is the same as in GR,
\begin{equation}
\label{eq:conservationequation}
    \nabla_{\mu}T^{\mu}{}_{\nu}=0.
\end{equation}

Note that Eq. \eqref{eq:cottonfieldequation} can be rewritten in terms of the Codazzi tensor \cite{Mantica:2022flg}, which satisfies the condition
\begin{equation}
\nabla_{\rho}\mathscr{C}_{\mu\nu}=\nabla_{\mu}\mathscr{C}_{\rho\nu},
\label{eq:codazzi}
\end{equation}
where
\begin{equation}
\label{eq:changecotton}
    \mathscr{C}_{\mu\nu}=R_{\mu\nu}-8\pi GT_{\mu\nu}-\frac{1}{6}(R-16\pi GT)g_{\mu\nu}.
\end{equation}
From this, we can observe that all solutions of Einstein’s equation, regardless of the presence of cosmological constant, are encompassed within Eq. \eqref{eq:cottonfieldequation}, when the Codazzi tensor 
is a constant. Thus, in Cotton gravity, the cosmological constant can be viewed as a mere integration constant or, in other words, as a manifestation of the gravitational effect.

We can express Eq. \eqref{eq:changecotton} in a different form \cite{Mantica:2022flg}
\begin{equation}
\label{eq:changecotton2}
    G_{\mu\nu}=8\pi GT_{\mu\nu}+\mathscr{C}_{\mu\nu}-\mathscr{C}g_{\mu\nu},
\end{equation}
where $\mathscr{C}=g^{\mu\nu}\mathscr{C}_{\mu\nu}$.
As is typical of many modified gravity theories, the deviation from GR can equally be modelled as due to some sort of modified matter sector. In this case, the terms involving the Codazzi tensor in Eq. \eqref{eq:changecotton2} can be interpreted as an anisotropic perfect fluid \cite{Gogberashvili:2023wed} in GR. 

Of course, there exist other intriguing, non-GR solutions to Eq. \eqref{eq:cottonfieldequation}. In a recent paper published by Harada \cite{Harada:2022edl}, a Schwarzschild-like solution is presented, which explains the rotation curve of galaxies through the distribution of baryons in Cotton gravity. Additionally, Harada analyzed the new Cotton gravity solution within the Solar system and constrained the associated parameters in a followed-up work \cite{Harada:2021bte}. Regardless of the theoretical debate surrounding the current unclear status of the theory concerning its predictability issue \cite{Clement:2023tyx,Sussman:2024iwk,Clement:2024pjl,Sussman:2024qsg}, it is important to further test the theory via observational data. To this end, we shall investigate cosmological perturbation in Cotton gravity and test its influence on the CMB and large scale structures.

\section{COSMOLOGICAL BACKGROUND AND PERTURBATIONS\label{cosmological background and perturbations}}

\subsection{\label{sec:level1}ISOTROPIC AND HOMOGENEOUS BACKGROUND}
Let us consider an isotropic and homogeneous cosmic background, which can be described by the Friedmann-Lemaître-Robertson-Walker (FLRW) metric,
\begin{equation}
ds^2=-dt^2+a^2(t)\left(\frac{dr^2}{1-Kr^2}+r^2d\Omega\right),
    \label{eq:FLRWmetric}
\end{equation}
where $K$ denotes the spatial curvature.

We assume that the distribution of matter is homogeneous throughout the universe, leading to the energy-momentum tensor of the form
\begin{equation}
\label{eq:energymomenttensor}
      T^{\mu}{}_{\nu}=
  \left(
\begin{matrix}
-\rho&0&0&0\\
0&p&0&0\\
0&0&p&0\\
0&0&0&p
\end{matrix}
\right).
\end{equation}
Unfortunately, upon substituting Eq. \eqref{eq:FLRWmetric} and Eq. \eqref{eq:energymomenttensor} into Eq. \eqref{eq:cottonfieldequation}, it reduces to the tautology ``0=0", which means that Cotton gravity does not allow us to distinguish between different spatial curvature. As shown in the paper \cite{Sussman:2023wiw,Sussman:2023eep,Mantica:2023ssd}, it is better to use the equivalent field equation, Eq. \eqref{eq:codazzi}, to investigate the evolution of the cosmic background under the cosmological principle. This spacetime can always contains a perfect fluid Codazzi tensor form displays a freedom in its parameters \cite{Mantica:2023ssd}, which is shown as 
\begin{equation}
    \mathscr{C}_{\mu\nu}=\mathscr{A}(t)u_{\mu}u_{\nu}+\mathscr{B}(t)g_{\mu\nu}+\frac{\Lambda}{3}g_{\mu\nu},
\end{equation}
where $\mathscr{A}$ and $\mathscr{B}$ are the arbitary function of the cosmic time constrained by these conditions, $\nabla_{\mu}\mathscr{A}=-\dot{\mathscr{A}}u_{\mu}$, $\nabla_{\mu}\mathscr{B}=-\dot{\mathscr{B}}u_{\mu}$, $\dot{\mathscr{B}}=-H\mathscr{A}$.
The evolution equation is given by:
\begin{equation}\label{friedmanneq}
      H^2=\frac{8\pi}{3}(\rho(a)+\Lambda)-\frac{K}{a^2}-\mathscr{B}(t).
\end{equation}
Again we emphasize that due to the aforementioned ``0=0'' problem, Cotton gravity allows non-unique evolution of the spatial curvature. In this article we simply assume that the background evolution in Cotton gravity is consistent with that in the $\Lambda$CDM model with zero spatial curvature for simplicity. We choose $\mathscr{B}=0$ and $K=0$. 
  
\subsection{\label{sec:COSMOLOGICAL PERTURBATION}COSMOLOGICAL PERTURBATION}
In the previous section, we have set the stage of our cosmological background evolution, which is the same as that in GR (GR solutions are also solutions to Cotton gravity). To understand the impact of modified gravity at the perturbation level, it is crucial to derive its perturbation theory, specifically the scalar perturbation, which contributes most of the temperature and $E$-mode polarization power spectrum in the CMB. In this study, we adopt the Newtonian gauge and concentrate on the scalar perturbation. In the calculation below, we use the natural unit system.

The expressions for the metric perturbation and the perturbation of the energy-momentum tensor are as follows:
\begin{widetext}
\begin{equation}
\label{eq:perturbationmetric}
      h_{\mu\nu}=
  \left(
\begin{matrix}
-2\Psi&0&0&0\\
0&a^2\cdot2\Phi&0&0\\
0&0&a^2\cdot2\Phi&0\\
0&0&0&a^2\cdot2\Phi
\end{matrix}
\right),
\end{equation}
\begin{equation}
\label{eq:perturbationenergymomenttensor}
      \delta T^{\mu}{}_{\nu}=
  \left(
\begin{matrix}
-\delta\rho&(\rho+p)\partial_1u&(\rho+p)\partial_2u&(\rho+p)\partial_3u\\
-\frac{(\rho+p)}{a^2}\partial_1u&\delta p+\partial_1^2\Pi&\partial_1\partial_2\Pi&\partial_1\partial_3\Pi\\
-\frac{(\rho+p)}{a^2}\partial_2u&\partial_2\partial_1\Pi&\delta p+\partial_2^2\Pi&\partial_2\partial_3\Pi\\
-\frac{(\rho+p)}{a^2}\partial_3u&\partial_3\partial_1\Pi&\partial_3\partial_2\Pi&\delta p+\partial_3^2\Pi
\end{matrix}
\right),
\end{equation}
where $u$ is the scalar velocity potential, and $\Pi$ is the anisotropic
inertia term. The field equation, Eq. \eqref{eq:cottonfieldequation}, becomes
    \begin{equation}
    \label{eq:perturbation1}
    \partial_i\left(u(3a\dot{a}(\rho+p)+3a^2(\dot{\rho}+\dot{p}))+3(\rho+p)a^2\dot{u}+3(\rho+p)a^2\Psi+3a^2\delta p+2a^2\delta\rho+4\nabla^{2}(\Phi-\Psi)+\nabla^2\Pi\right)=0,
\end{equation}
\begin{equation}
    \label{eq:perturbation2}
\partial_i\partial_j\left(-\frac{1}{2}(\rho+p)u+\frac{\dot{a}}{a}(\Phi-\Psi)+(\dot{\Phi}-\dot{\Psi})-\frac{1}{2}a(\dot{a}\Pi+a\dot{\Pi})\right)=0,
\end{equation}
\begin{equation}
    \label{eq:perturbation3}
\partial_i\left(3a\dot{a}(\rho+p)u-a^2\delta\rho+2\nabla^{2}(\Psi-\Phi)+a^2\nabla^2\Pi\right)=0,
\end{equation}
\begin{eqnarray}
    \label{eq:perturbation4}
&&\qquad \qquad \quad 3a\dot{a}(\delta p+\delta\rho)+a^2\dot{\delta\rho}(6a(p+\rho)\dot{a}+2a^2\dot{\rho})\Phi)+3(\rho+p)\partial_i\partial_ju\nonumber \\ 
&&+3a^2(\rho+p)\dot{\Phi}+\frac{\dot{a}}{a}(2\nabla^2(\Phi-\Psi)-6\partial_i\partial_i(\Phi-\Psi))+2\nabla^2(\dot{\Phi}-\dot{\Psi})-6\partial_i\partial_i(\dot{\Phi}-\dot{\Psi})=0.
\end{eqnarray}
\end{widetext}
The conservation law in Eq. \eqref{eq:conservationequation} becomes
\begin{equation}
    \label{eq:perturbation5}
\delta p+\nabla^2\Pi+(\rho+p)\dot{u}+\dot{p}u+(\rho+p)\Psi=0,
\end{equation}
\begin{equation}
\label{eq:perturbation6}
\dot{\delta\rho}+\frac{3\dot{a}}{a}(\delta\rho+\delta p)+\nabla^2\left(\frac{\rho+p}{a^2}u+\frac{\dot{a}}{a}\Pi\right)+3(\rho+p)\dot{\Phi}=0.
\end{equation}
Eq. \eqref{eq:perturbation1} and Eq. \eqref{eq:perturbation4} can be derived from the other four equations. For simplicity, we shall rewrite the four equations in their Fourier representation:
\begin{equation}
\label{eq:1}
    -\frac{1}{2}(\rho+p)u+\frac{\dot{a}}{a}(\Phi-\Psi)+(\dot{\Phi}-\dot{\Psi})-\frac{1}{2}a(\dot{a}\Pi+a\dot{\Pi})=0,
\end{equation}
\begin{equation}
    \label{eq:2}
3a\dot{a}(\rho+p)u-a^2\delta\rho-2k^{2}(\Psi-\Phi)-a^2k^2\Pi=0,
\end{equation}
\begin{equation}
    \label{eq:3}
\delta p-k^2\Pi+(\rho+p)\dot{u}+\dot{p}u+(\rho+p)\Psi=0,
\end{equation}
\begin{equation}
\label{eq:4}
\dot{\delta\rho}+\frac{3\dot{a}}{a}(\delta\rho+\delta p)-k^2\left(\frac{\rho+p}{a^2}u+\frac{\dot{a}}{a}\Pi\right)+3(\rho+p)\dot{\Phi}=0.
\end{equation}
From Eq. \eqref{eq:1}-\eqref{eq:4}, we can derive the relationship between the two gravitational potential $\Psi$ and $\Phi$, which is given by
\begin{equation}
\label{eq:5}
     \dot{a}(\Phi+\Psi)-a(\dot{\Phi}+\dot{\Psi})+\frac{1}{2}a(a\dot{\Pi}+\dot{a}\Pi)=0.
\end{equation}
It should be emphasized again that we only derive the linear perturbation theory under the assumption that the background evolution is the same as that in GR, with the Friedmann equation unmodified. Otherwise the perturbation theory would depend on $\mathscr{K}$ and $\gamma$ in Eq.\eqref{eq:FLRWmetric}.
From Eq. \eqref{eq:5} we can obtain
\begin{equation}
    \Phi+\Psi=\beta(k)\times a-\frac{1}{2}a^2\Pi,
    \label{eq:111}
\end{equation}
where $\beta$ is a function of the wave number $k$, which is the Cotton gravity parameter. For simplicity, we shall restrict to the simplest case that $\beta$ is a constant.

Observe that the perturbation functions of Cotton gravity, except for the traceless part, remain the same as those of GR. By redefining the perturbation energy-momentum tensor as $\tilde{\delta\rho}=\delta\rho+\frac{2k^2\beta}{a}$, $\tilde{\delta p}=\delta p-\frac{2k^2\beta}{a}$, $\tilde{\Pi}=\Pi-\frac{2\beta}{a}$, all the perturbation equations become the same as those in GR. Thus, at the first perturbation order level, the effects of Cotton gravity can be interpreted as an anisotropic fluid, as mentioned earlier. Consequently, the curvature perturbation does not evolve when the perturbation extends beyond the Hubble horizon. Additionally, we do not consider the influence of Cotton gravity on inflation. Hence, we have established that the initial conditions are the same as those in the standard cosmology model. What then is the difference between Cotton gravity and GR in cosmological perturbation?

In the $\Lambda$CDM model, the anisotropic stress is mainly contributed by the high-order moments of relativistic particles, such as photons and neutrinos, during the radiation domination era. However, after decoupling and during the matter domination era, the anisotropic stress decreases and becomes negligible. Therefore, during this period, we can assume that the two gravitational potentials remain constant while the matter density perturbation increases linearly with $a$. In Cotton gravity, however, the case is different. If the best-fit value of $\beta$ is positive, this additional term will cause the gravitational potential to increase during the matter domination era. Consequently, it contributes to the growth of structures, leading to an increase in the matter power spectrum and affecting the Integrated Sachs-Wolfe (ISW) effect in both the early and late times.

On the other hand, as already mentioned, at the first-order perturbation level, Cotton gravity is identical to GR except for the traceless part. This means that the sound horizon remains the same as in $\Lambda$CDM model when other cosmological parameters are held unchanged. Therefore, it can be anticipated that Cotton gravity will have minimal influence on the position of acoustic peaks.

\section{CMB CONSTRAINTS ON THE PARAMETERS\label{CMB constraints on the parameters}}
We utilize the public Einstein-Boltzmann solver, MGCLASS II \cite{Sakr:2021ylx}, which is a modification of the publicly available CLASS \cite{Blas:2011rf} code. We further adapt the code to incorporate the perturbation functions in Eq. \eqref{eq:1}-\eqref{eq:4}. There will be more detail in Appendix \ref{appendix}. This adaptation allows us to accurately calculate the CMB power spectrum based on the given cosmological parameters and Cotton gravity parameter.

\subsection{\label{INFLUENCE OF COTTON GRAVITY PARAMETER}INFLUENCE OF COTTON GRAVITY PARAMETER}
Cotton gravity will influence the evolution of perturbations, especially the CMB and large scale structures. We will discuss these separately.

\begin{figure}[htbp]
\includegraphics[scale=0.35]{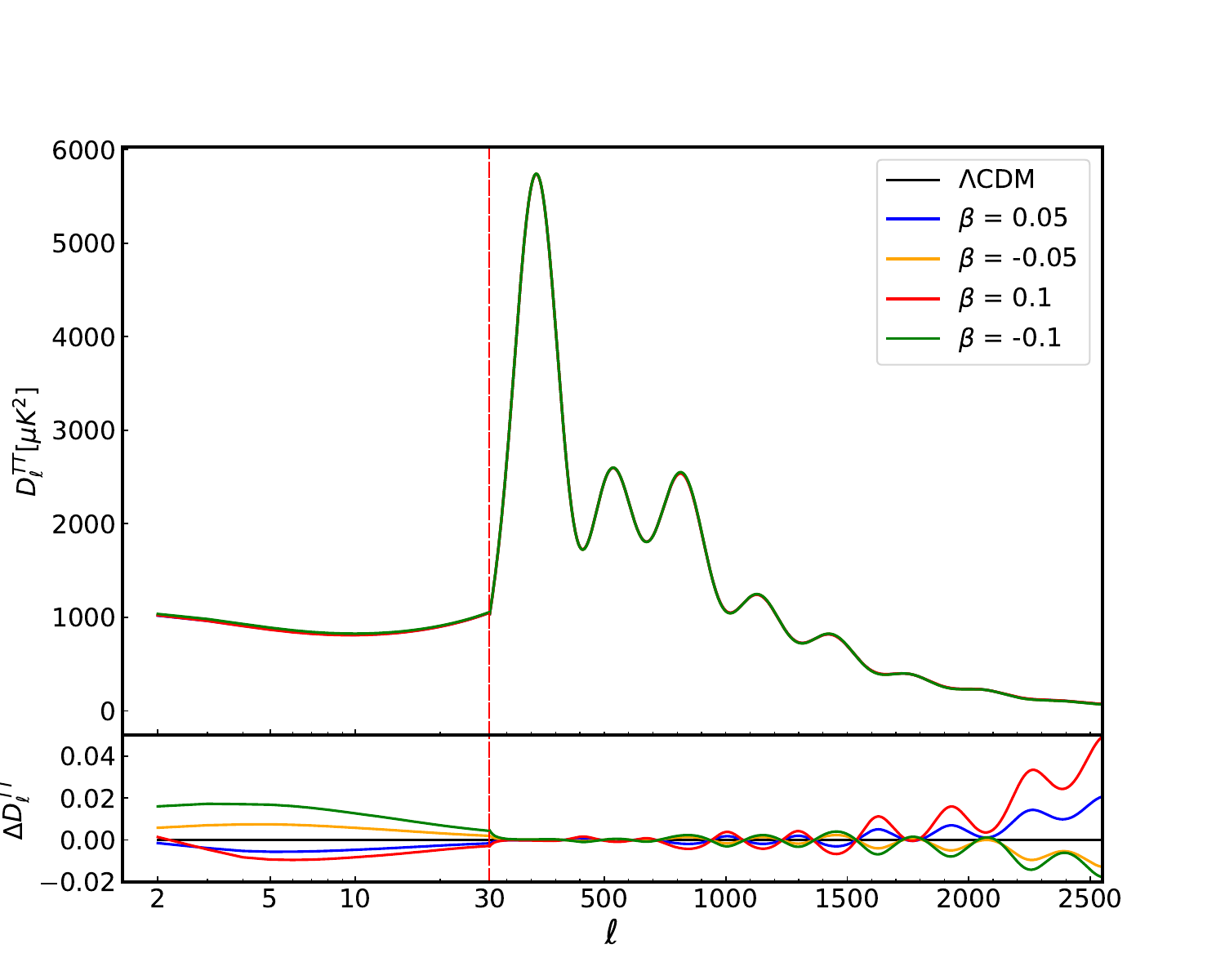}
\caption{The CMB temperature power spectra produced by different $\beta$, where the set of its value is $\left\{-0.1, -0.05, 0, 0.05, 0.1\right\}$ ($\Lambda$CDM corresponds to $\beta=0$). Their differences with CMB temperature power spectrum in $\Lambda$CDM model is shown in the lower panel.}
\label{fig:diffparameter0}
\end{figure}
Fig.~\ref{fig:diffparameter0} displays the CMB temperature power spectra for different values of $\beta$, along with their difference compared to the $\Lambda$CDM model. Here,
\begin{equation}
    D^{TT}_l=\frac{l(l+1)C^{TT}_l}{2\pi},
\end{equation}
and
\begin{equation}
    \Delta D^{XX}_l=(D^{XX}_l-D^{XX}_{l,\Lambda CDM})/D^{XX}_{l,\Lambda CDM},
\end{equation}
where $X$ represents $T$, $E$ or $\phi$, corresponding to the CMB temperature, $E$-mode polarization and lensing potential, respectively.
The values of $\beta$ used in Fig.~\ref{fig:diffparameter0} are $-0.1, -0.05, 0, 0.05,$ and $0.1$, respectively. It is important to note that other parameters have the same values as the six cosmological parameters in the $\Lambda$CDM model.

Looking at Fig.~\ref{fig:diffparameter0}, we can observe a significant influence of the parameter $\beta$ on the temperature power spectrum in both low-$\ell$ and high-$\ell$ regions. In the low-$\ell$ regions, the spectrum is enhanced due to the effect of Cotton gravity on the ISW effect. Nevertheless, in the high-$\ell$ regions, the difference of CMB temperature power spectrum between Cotton gravity and $\Lambda$CDM model is mainly attributed to the lensing effect. If we subtract the lensing effect, the CMB temperature and $E$-mode polarization power spectrum are hardly influenced by Cotton gravity in the high-$\ell$ region. It is worth noting that when $\beta$ is small, Cotton gravity has opposite effects depending on whether the Cotton parameter $\beta$ is positive or negative.

\begin{figure}[htbp]
\includegraphics[scale=0.35]{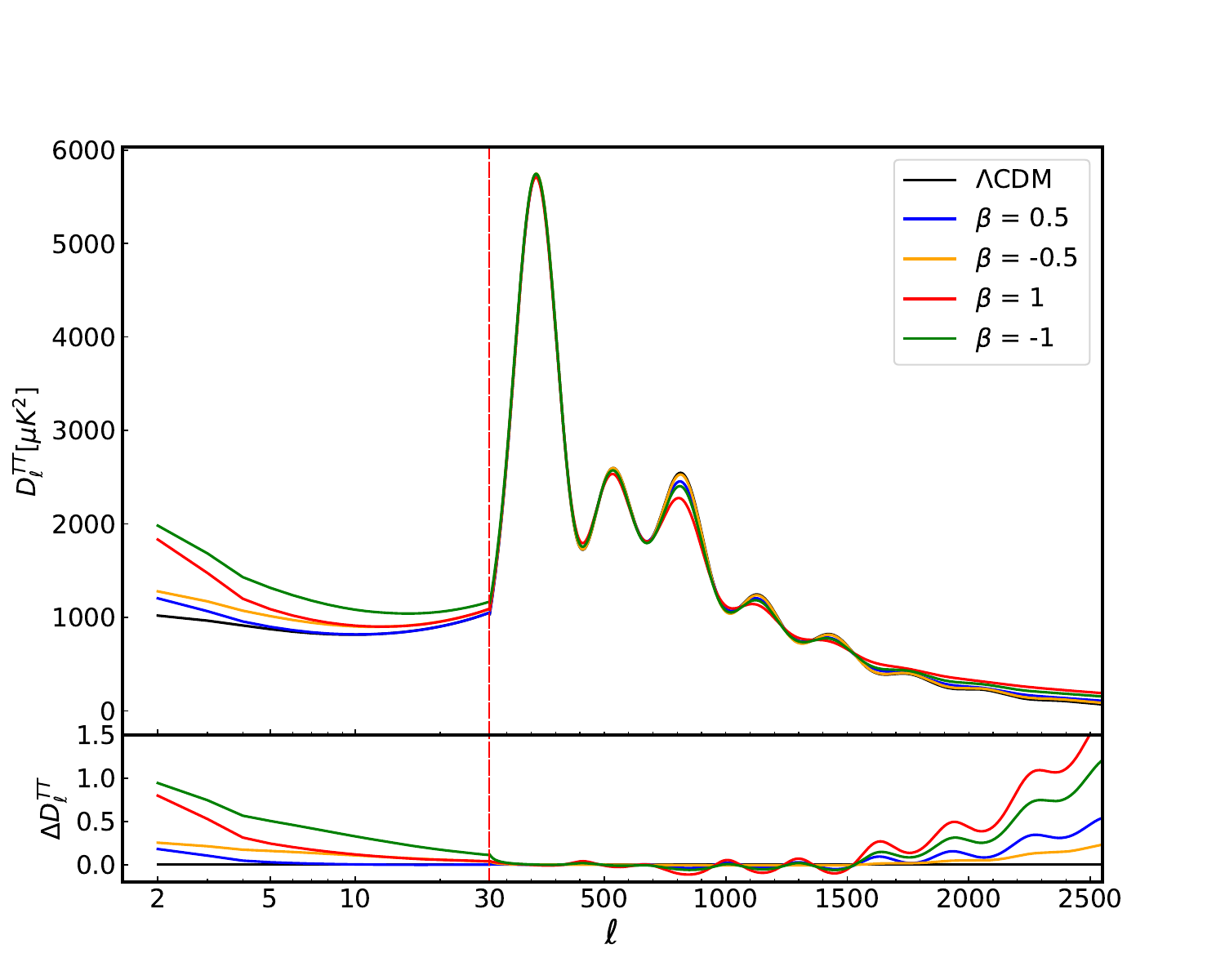}
\caption{The CMB temperature power spectra produced by $\beta \in\left\{-1, -0.5, 0, 0.5, 1\right\}$. Their difference with CMB temperature power spectrum in $\Lambda$CDM model ($\beta=0$) is shown in the lower panel.}
\label{fig:diffparameter1}
\end{figure}

However, the situation changes when the parameter $\beta$ takes a large value, as demonstrated in Fig.~\ref{fig:diffparameter1}, where the values of $\beta$ used are now $-1, -0.5, 0, 0.5, 1$. In both low-$\ell$ and high-$\ell$ regions, the effect of Cotton gravity becomes more significant compared to the original contribution. Therefore, for both positive and negative values of $\beta$, the CMB temperature power spectrum is enhanced in the low-$\ell$ and high-$\ell$ regions, which is attributed to the same reasons discussed above.

\begin{figure}[htbp]
\includegraphics[scale=0.30]{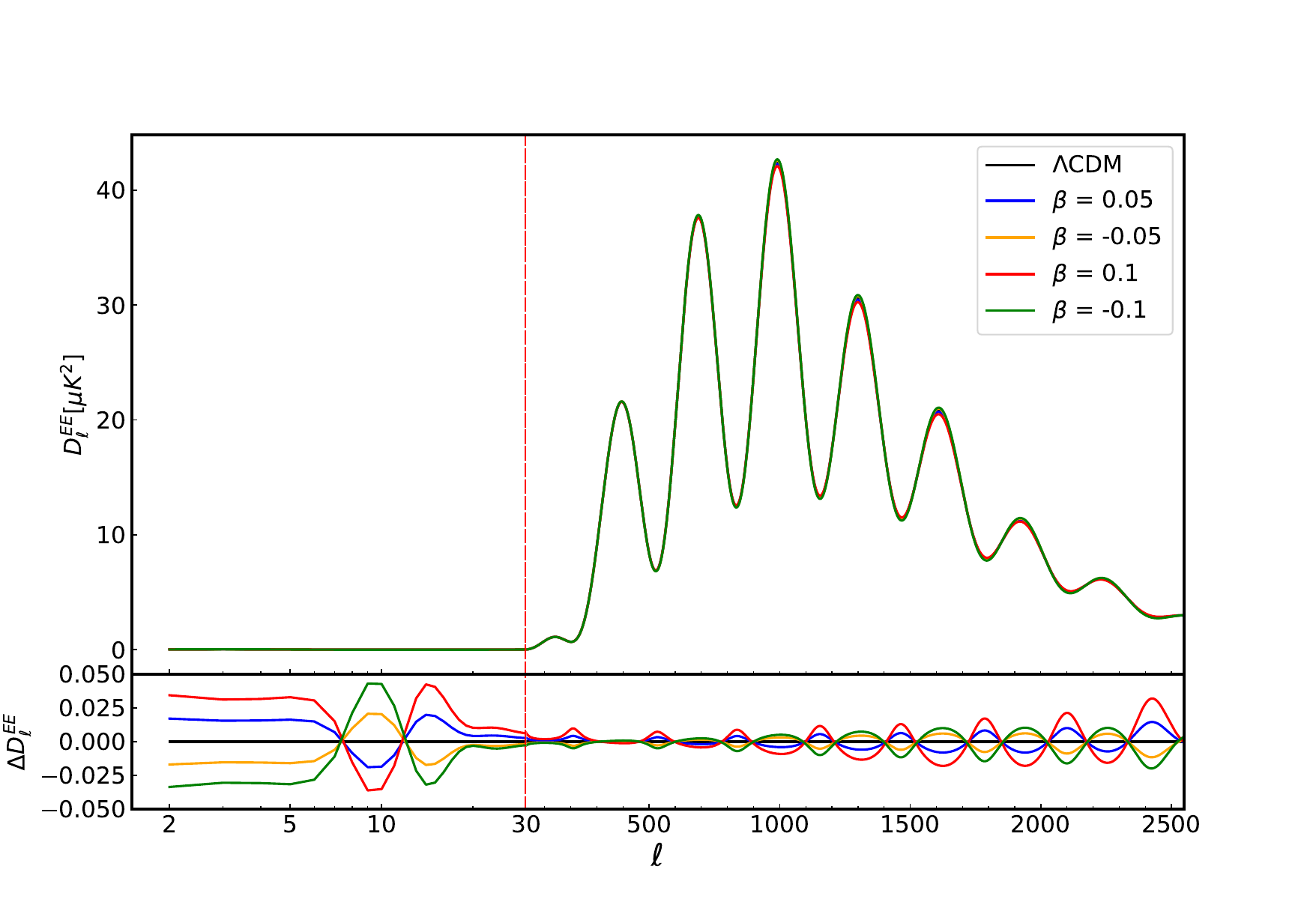}
\caption{The CMB $E$-mode polarization power spectra produced by $\beta \in \left\{-0.1, -0.05, 0, 0.05, 0.1\right\}$. Their difference with CMB $E$-mode polarization power spectrum in $\Lambda$CDM model ($\beta=0$) is shown in the lower panel.}
\label{fig:diffparameter2}
\end{figure} 

\begin{figure}[htbp]
\includegraphics[scale=0.33]{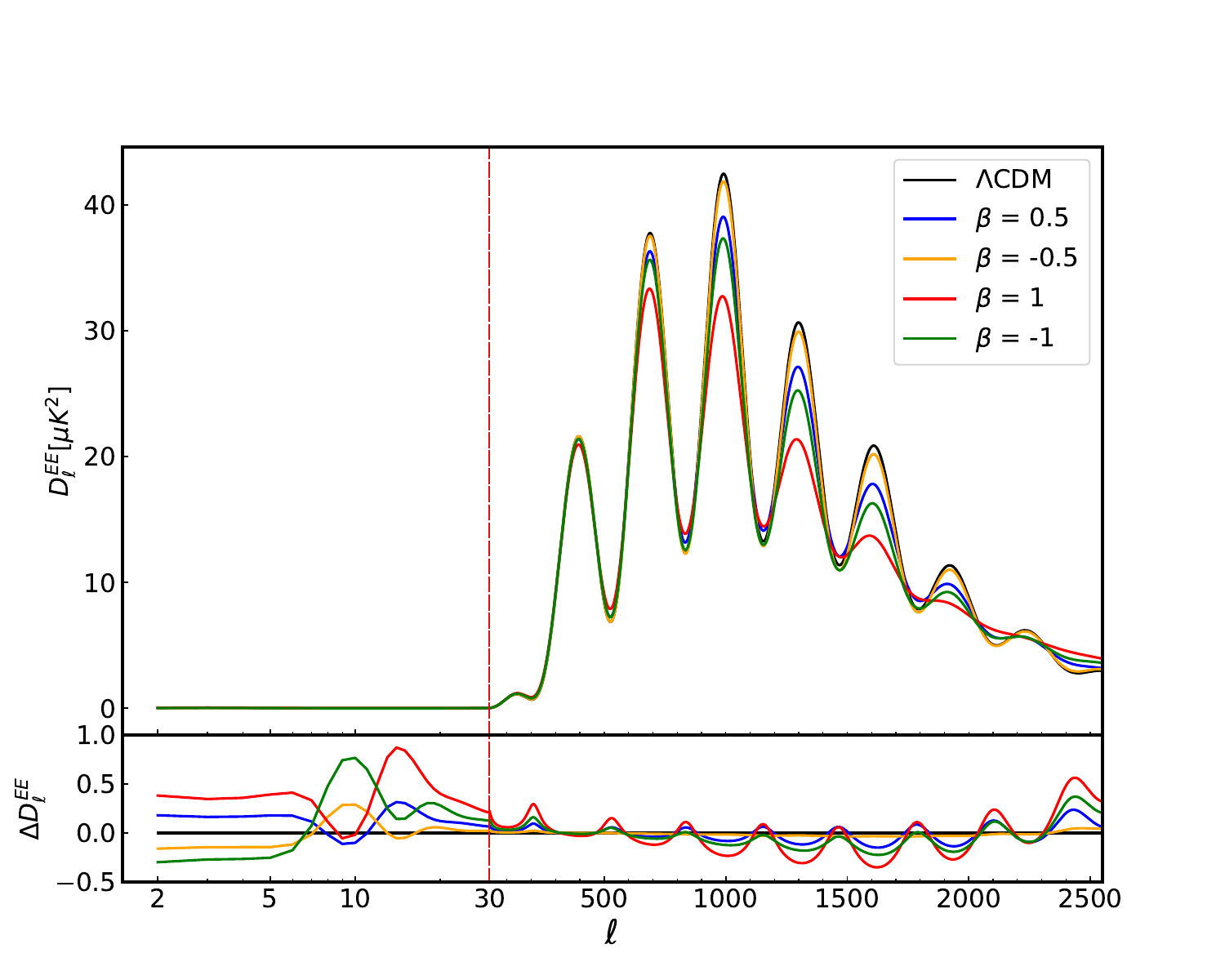}
\caption{The CMB $E$-mode polarization power spectra produces by $\beta \in \left\{-1, -0.5, 0, 0.5, 1\right\}$. Their difference with CMB $E$-mode polarization power spectrum in $\Lambda$CDM model ($\beta=0$)  is shown in the lower panel.}
\label{fig:diffparameter3}
\end{figure} 

Fig.~\ref{fig:diffparameter2} and Fig.~\ref{fig:diffparameter3} show the $E$-mode polarization power spectra of different $\beta$ parameters. In the low-$\ell$ region, the polarization signal from reionization is significant. In Cotton gravity, the gravitational potential varies with different $\beta$ values, affecting the reionization polarization signal. Consequently, the $E$-mode polarization power spectrum in Cotton gravity is modified. In the high-$\ell$ region, the influence of Cotton gravity is basically the same as that on the temperature power spectrum.

\begin{figure}[htbp]
\includegraphics[scale=0.35]{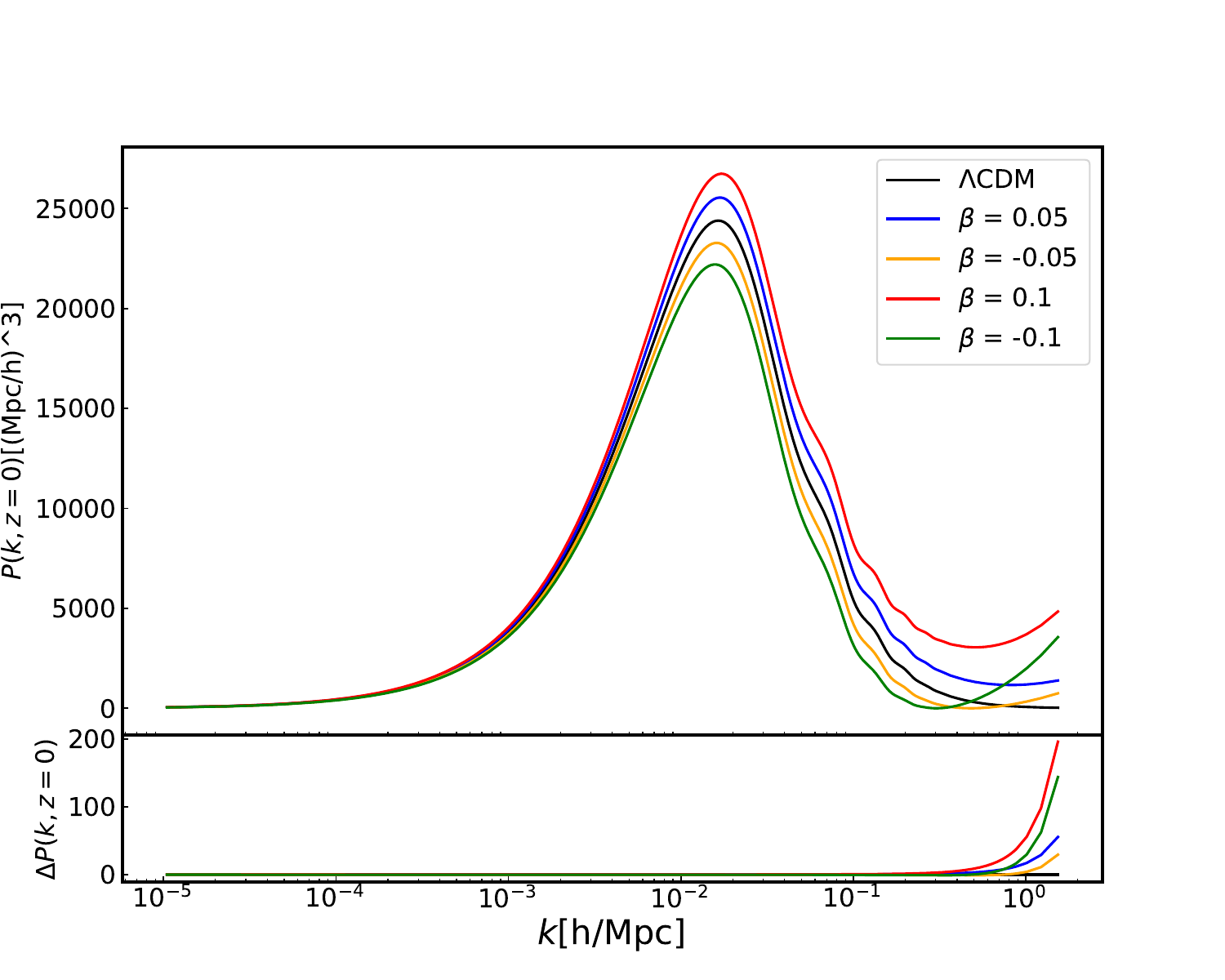}
\caption{The matter power spectra corresponding to Cotton parameter $\beta \in \left\{-0.1, -0.05, 0, 0.05, 0.1\right\}$ and Their difference with matter power spectrum in $\Lambda$CDM model ($\beta=0$)  is shown in the lower panel.}.
\label{fig:PS_diffeerent}
\end{figure}

Moreover, the matter power spectrum is also influenced by Cotton gravity, as shown in Fig.~\ref{fig:PS_diffeerent}. Qualitatively, we can observe that Cotton gravity has a larger impact on the matter power spectrum compared to the CMB power spectrum, especially at small scale. As mentioned earlier, when the parameter $\beta$ is positive, it enhances the gravitational potential and leads to an increase in the matter power spectrum. Conversely, when $\beta$ is negative, the effect is reversed.

\begin{figure}[htbp]
\includegraphics[scale=0.35]{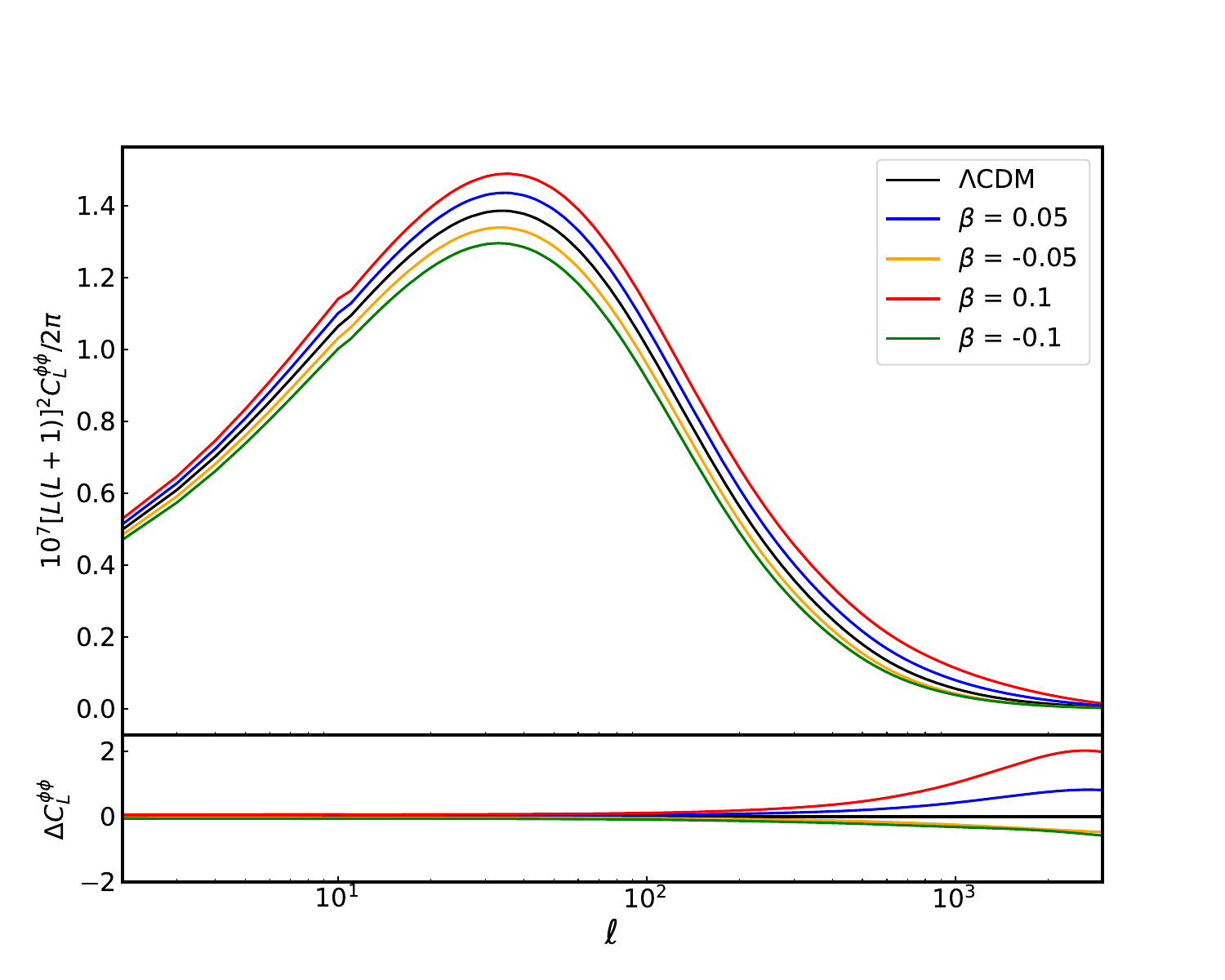}
\caption{CMB lensing-potential power spectra produced by $\beta \in \left\{-0.1, -0.05, 0, 0.05, 0.1\right\}$. Their difference with CMB lensing-potential power spectrum in $\Lambda$CDM model ($\beta=0$)  is shown in the lower panel.}
\label{fig:diffparameter1_lensing}
\end{figure}

Due to the different gravitational potential and matter power spectrum, the lensing effect in CMB is also modified in Cotton gravity. In Fig.~\ref{fig:diffparameter1_lensing}, it can be seen that the CMB lensing-potential power spectrum is dramatically impacted by the effect of Cotton gravity parameter.

\subsection{\label{BESTFIT PARAMMETERS}BEST-FIT PARAMETERS}
To obtain the best-fit parameter for Cotton gravity, we utilize the MCMC method \cite{Brinckmann:2018cvx, Audren:2012wb} with the Planck 2018 data, which includes high-$\ell$ TT, TE, EE, low-$\ell$ EE, low-$\ell$ TT and lensing likelihood \cite{Planck:2018vyg, Planck:2019nip}. More specifically, we use the data from $\texttt{COM\_Likelihood\_Data-baseline\_R3.00}$, which is publicly available on the Planck website \footnote{\url{https://pla.esac.esa.int/\#cosmology}}. In addition to varying the parameter $\beta$, we also incorporate other cosmological parameters such as baryon density $\Omega_{\mathrm{b}}h^2$, dark matter density $\Omega_{\mathrm{c}}h^2$, angular acoustic scale $100\theta_{*}$, optical depth $\tau$, primordial comoving curvature power spectrum amplitude $\ln(10^{10}A_{\mathrm{s}})$, and scalar spectral index $n_\mathrm{s}$, during the MCMC analysis. And We have imposed flat priors on this parameters. Table \ref{tab:table0} shows their prior range.
\begin{table}[htbp]
\centering
\begin{tabular}{|c|c|}
\hline
\textrm{}&
\textrm{Prior range \qquad}\\
\colrule
$\beta$&[-1, 1]\\
$\Omega_{\mathrm{b}}h^2$&[0.018, 0.03]\\
$\Omega_{\mathrm{c}}h^2$&[0.1, 0.2]\\
$100\theta_{*}$&[1.03, 1.05]\\
$\tau$&[0.004, 0.12]\\
$\ln(10^{10}A_{\mathrm{s}})$&[2.7, 3.5]\\
$n_\mathrm{s}$&[0.9, 1.1]\\
\hline
\end{tabular}
\caption{Prior ranges for Cotton gravity MCMC analysis.}
\label{tab:table0}
\end{table}

\begin{table}[htbp]
\begin{ruledtabular}
\begin{tabular}{c|c|c}
\textrm{}&
\textrm{Cotton gravity \qquad}& 
\textrm{$\Lambda$CDM \cite{Planck:2018vyg}}\\
\colrule
$\beta$&0.110$\pm$0.010&\\
$\Omega_{\mathrm{b}}h^2$&0.02239$\pm$0.00016 &0.02237$\pm$0.00015\\
$\Omega_{\mathrm{c}}h^2$&0.1194$\pm$0.0014&0.1200$\pm$0.0012\\
$100\theta_{*}$&1.04212$\pm$0.00031&$1.04110\pm$0.00031\\
$\tau$&0.0549$\pm$0.0078&0.0544$\pm$0.0073\\
$\ln(10^{10}A_{\mathrm{s}})$&3.046$\pm$0.015&3.044$\pm$0.014\\
$n_\mathrm{s}$&0.9686$\pm$0.0046&0.9649$\pm$0.0042\\
$\chi^2$&1388&1385
\end{tabular}
\end{ruledtabular}
\caption{1-$\sigma$ confidence interval of the parameters in Cotton gravity and in $\Lambda$CDM model.}
\label{tab:table1}
\end{table}

\begin{figure}[htbp]
\includegraphics[scale=0.35]{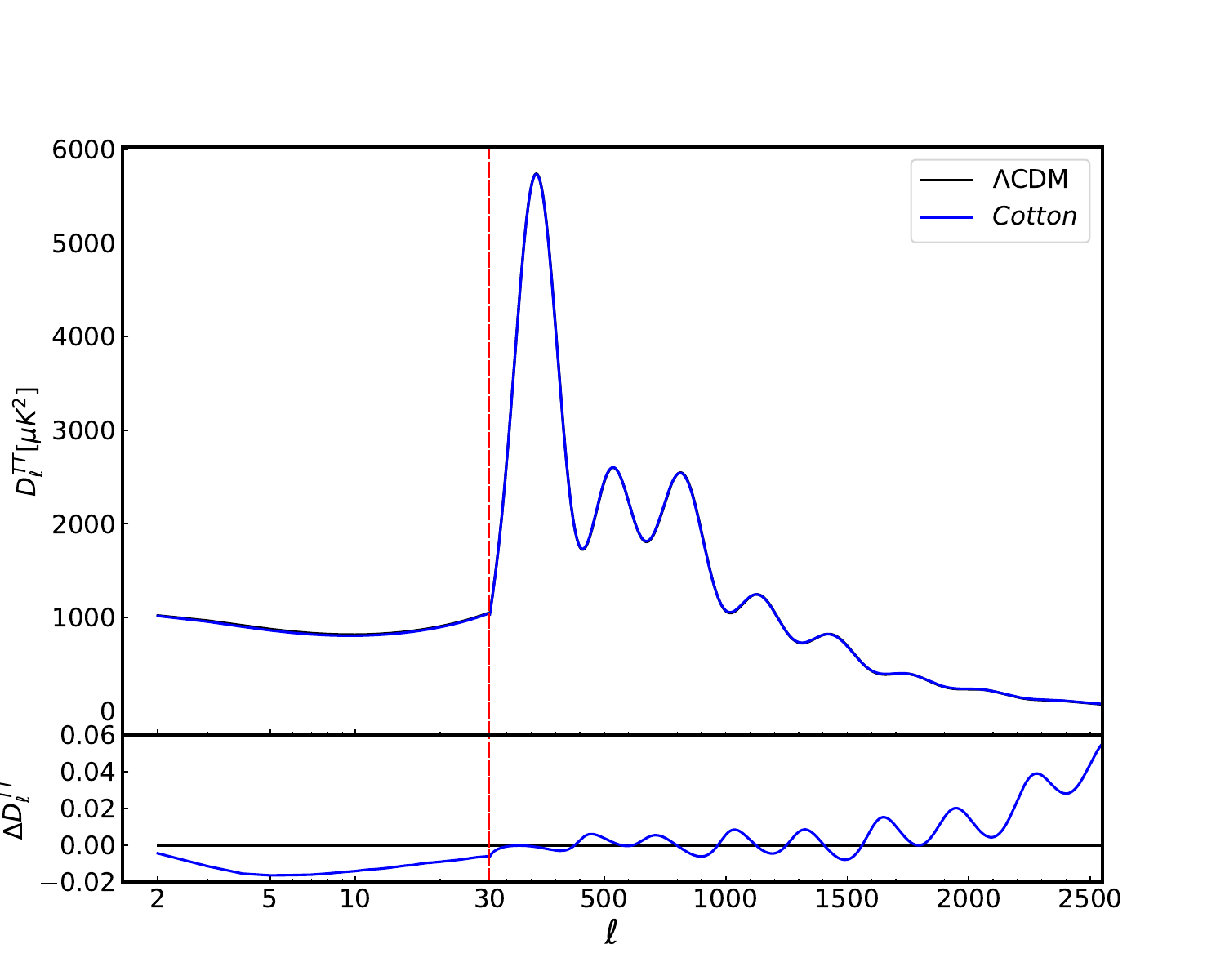}
\caption{The CMB temperature power spectra produced by Cotton gravity and $\Lambda$CDM model both with their best-fit parameters. The difference of them is shown in the lower panel.}
\label{fig:cotton+c_c_tt}
\end{figure}

\begin{figure}[htbp]
\includegraphics[scale=0.35]{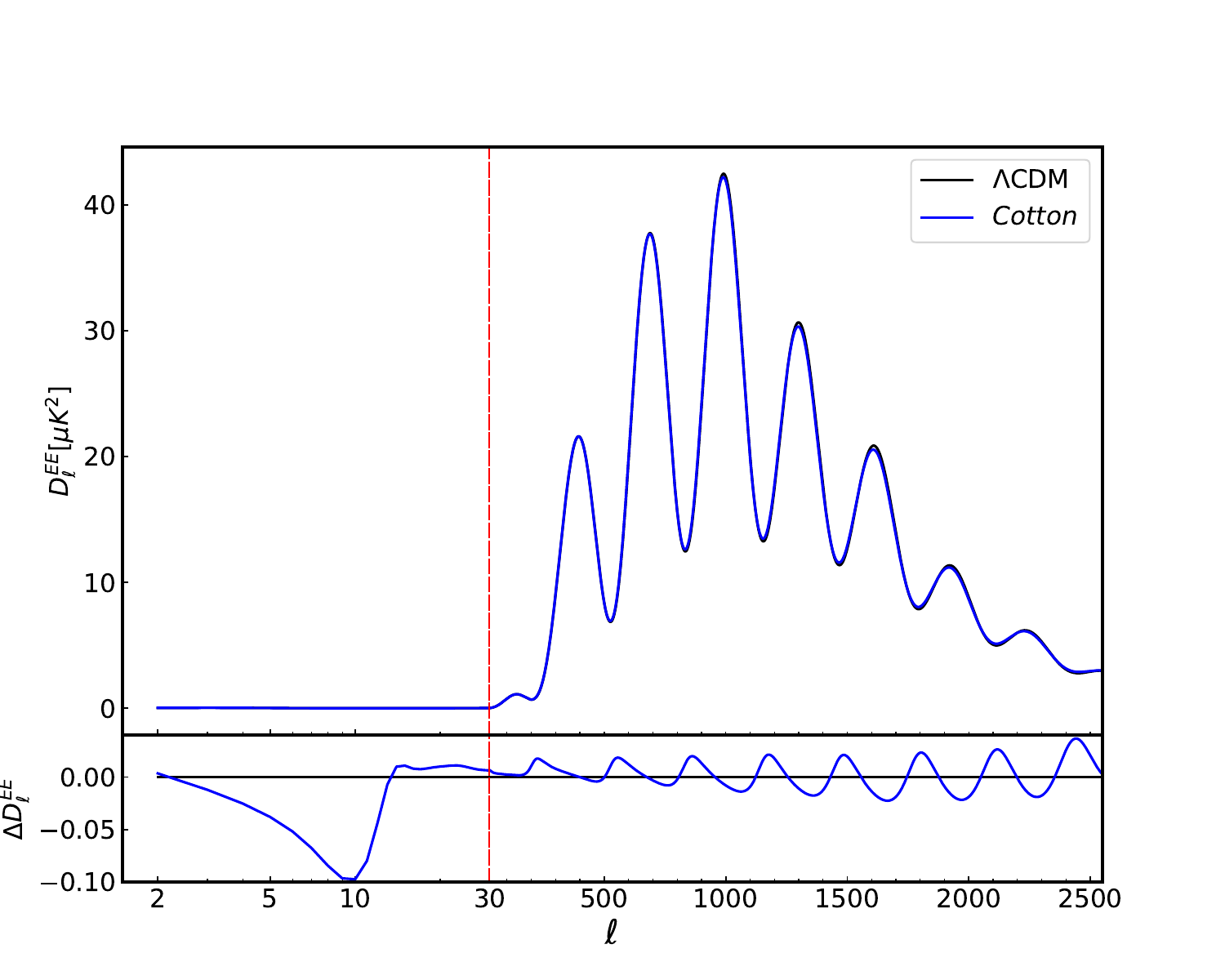}
\caption{The CMB $E$-mode polarization power spectra produced by Cotton gravity and $\Lambda$CDM model both with their best-fit parameters. The difference of them is shown in the lower panel.}
\label{fig:cotton+c_c_ee}
\end{figure}

Table \ref{tab:table1} presents the  1-$\sigma$ confidence intervals for the seven parameters in both Cotton gravity and $\Lambda$CDM model. Generally, all parameters in Cotton gravity model, excluding the model parameter $\beta$, exhibit little difference compared to those in $\Lambda$CDM model. Additionally, the $\chi^2$ difference between the two models is negligible. Furthermore, in Fig.~\ref{fig:cotton+c_c_tt} and Fig.~\ref{fig:cotton+c_c_ee}, we illustrate that the differences in the CMB temperature and the $E$-mode power spectrum calculated using the best-fit parameters for both models are very small.
The MCMC simulation reveals that the best-fit value of $\beta$ is nonzero, but rather
\begin{equation}
    \beta=0.110^{+0.011}_{-0.010},
\end{equation}
within the 1-$\sigma$ confidence interval. This suggests that, when considering Cotton gravity with the other six cosmological parameters, CMB data does favor a nonzero value of $\beta$.

\begin{figure}[htbp]
\includegraphics[scale=0.35]{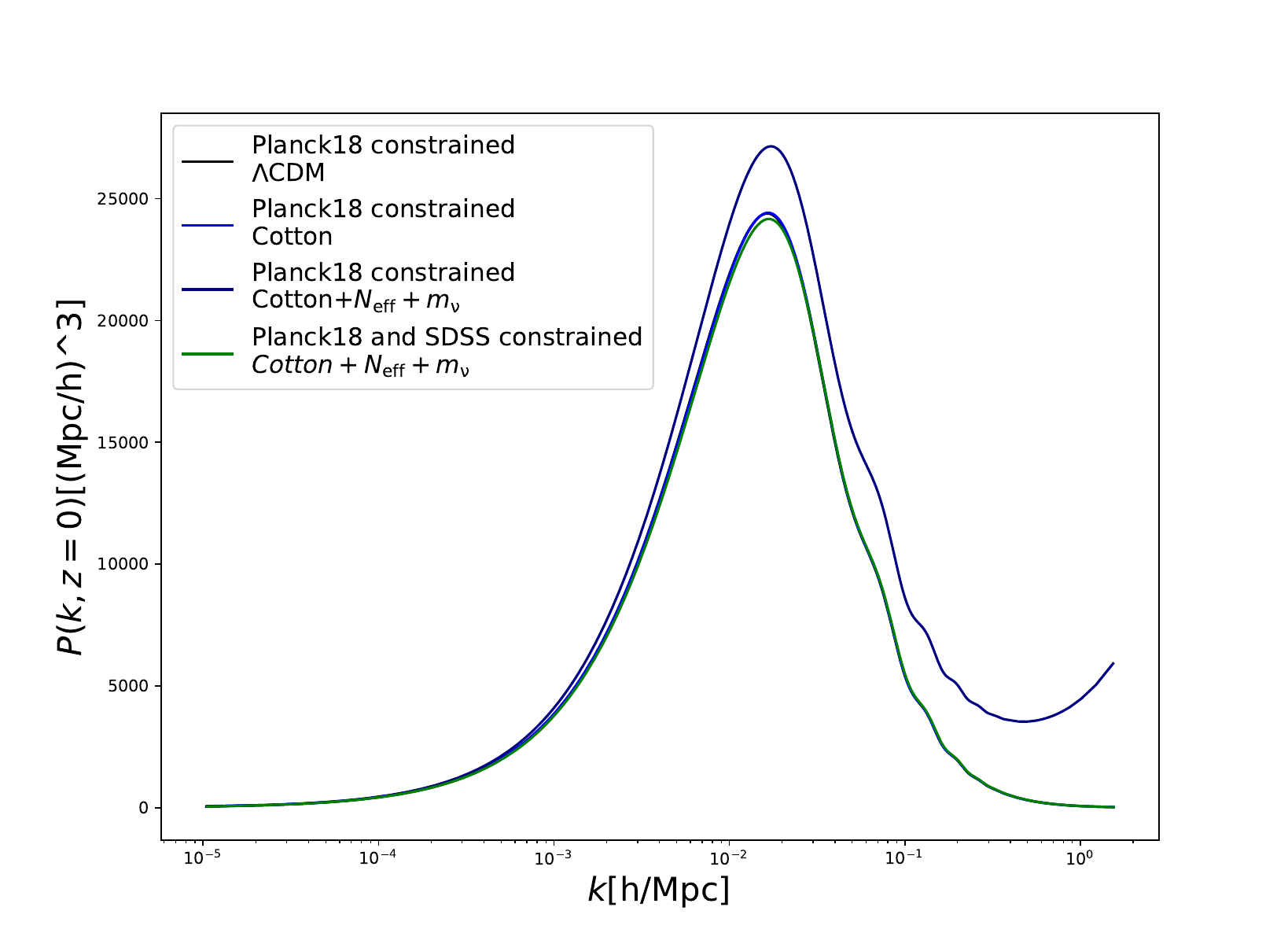}
\caption{The difference between matter power spectrum predicted by Cotton gravity, Cotton gravity including massive neutrinos, and $\Lambda$CDM model.}
\label{fig:PS1}
\end{figure}

However, it is important to note that the influence of the Cotton gravity parameter extends beyond its effect on the CMB data. Specifically, it also impacts the matter power spectrum. As shown in Fig.~\eqref{fig:PS_diffeerent} and Fig.~\eqref{fig:diffparameter1_lensing}, the matter power spectrum and CMB lensing-potential power spectrum deviate significantly from that predicted by the $\Lambda$CDM when $\beta$ becomes 0.1. Now, when $\beta$ takes a value close to its best fit 0.11, which can be considered relatively large, the disparities in matter power spectra are visually depicted in Fig.~\ref{fig:PS1}, highlighting the contrasting behavior between the two models, especially at small scale.
Such disparities lead us to the question whether the non-zero Cotton gravity parameter $\beta$ is even viable. 

Nevertheless, it is essential to understand why the CMB data favors a non-zero $\beta$. Considering the theory from another perspective, the effect of Cotton gravity may exhibit degeneracy with other physical processes that influence the anisotropic stress. Within the $\Lambda$CDM model, the anisotropic stress primarily arises from the high-order moments of relativistic particles such as photons and neutrinos during the radiation domination period. Therefore, the influence of relativistic particles on the anisotropic stress must be related to the best-fit value of $\beta$. To study it, we introduce the neutrino model which has three massive neutrinos with degenerate mass. This model has two parameters: the effective number of relativistic species $N_{\mathrm{eff}}$ and the mass of one neutrino $m_{\mathrm{\nu}}$. The total neutrino mass $\sum m_{\mathrm{\nu}}$ is the product of $N_{\mathrm{eff}}$ and $m_{\mathrm{\nu}}$. These two parameters are taken into consideration during the MCMC analysis. We will impose flats prior on $N_{\mathrm{eff}}$ and $m_{\mathrm{\nu}}$ over the range 0$<N_{\mathrm{eff}}<$1 and $m_{\mathrm{\nu}}>$0. 
\begin{table}[htbp]
\begin{ruledtabular}
\begin{tabular}{c|c|c}
\textrm{}&
\textrm{Cotton gravity extension}& 
\textrm{$\Lambda$CDM + $N_{\mathrm{eff}}$ + $m_{\mathrm{\nu}}$}\\
\colrule
$\beta$ & -0.0002$\pm$0.0019 &\\
$\Omega_{\mathrm{b}}h^2$ & 0.02224$\pm$0.00024 &0.02225$\pm$0.00022\\
$\Omega_{\mathrm{c}}h^2$ & 0.1161$\pm$0.0030 &0.1163$\pm$0.0028\\
$100\theta_{*}$&1.04235$\pm$0.00052& 1.04240$\pm$0.00051\\
$\tau$&0.0524$\pm$0.0077&0.0520$\pm$0.0077\\
$\ln(10^{10}A_{\mathrm{s}})$&3.031$\pm$0.018&3.028$\pm$0.018\\
$n_{\mathrm{s}}$&0.9549$\pm$0.0089& 0.9620$\pm$0.0086\\
$\sum m_{\nu}$ & $\sum m_{\nu}\text{\textless}$0.28 (95$\%$) & $\sum m_{\nu}\text{\textless}$0.26 (95$\%$)\\
$N_{\mathrm{eff}}$ & 2.73$\pm$0.38 (95$\%$) & 2.77$\pm$0.36 (95$\%$)\\
$\chi^2$&1389&1389
\end{tabular}
\end{ruledtabular}
\caption{1-$\sigma$ confidence interval of the parameters in the Cotton gravity with neutrino, compared to that in the $\Lambda$CDM model.}
\label{tab:table2}
\end{table}

\begin{figure}[htbp]
\includegraphics[scale=0.35]{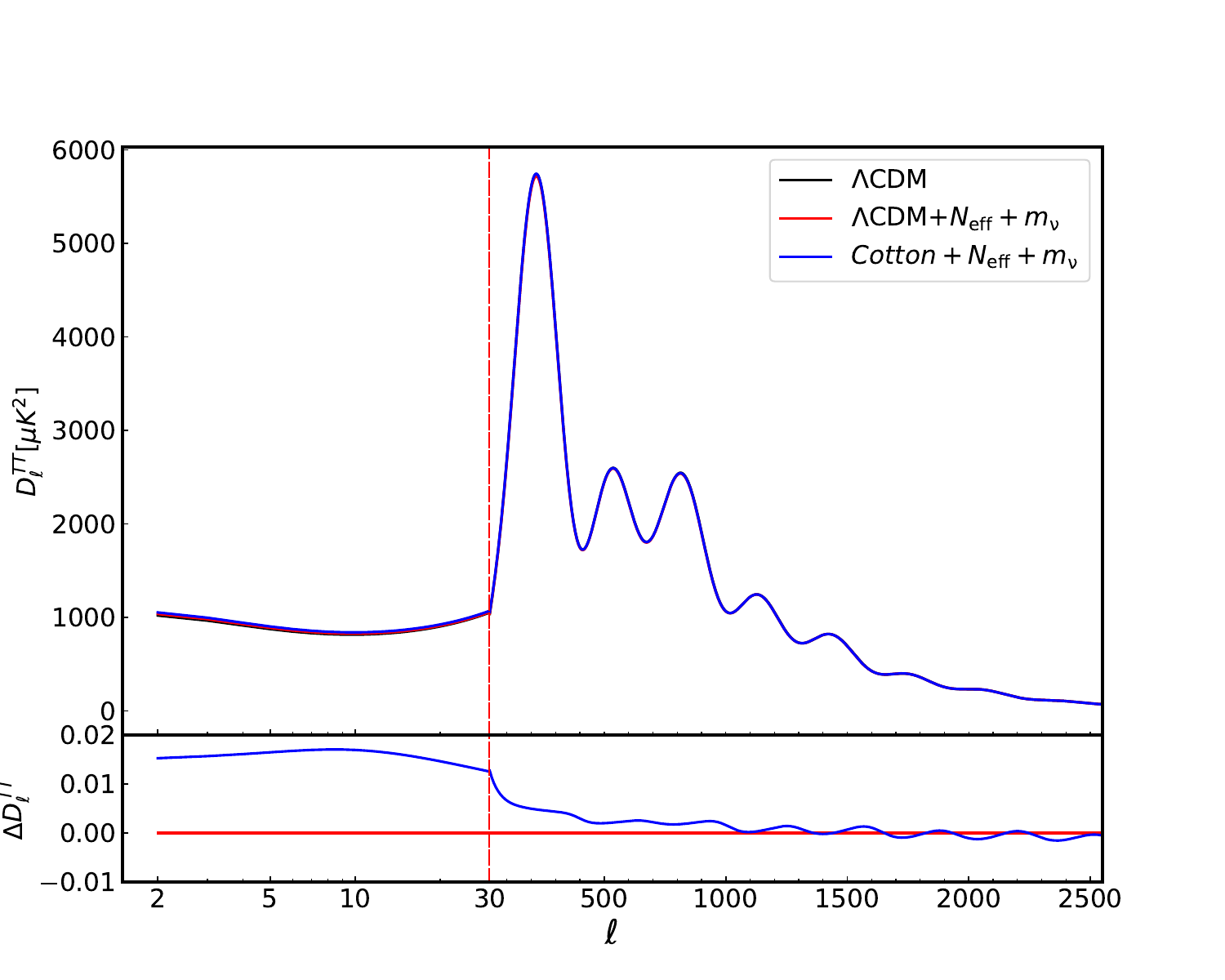}
\caption{The CMB temperature power spectrum of $\Lambda$CDM, $\Lambda$CDM extension and Cotton gravity extension with their best-fit parameters and the difference between the latter two is shown in the lower panel.}
\label{fig:Cotton+neutrino_c_tt}
\end{figure}
\begin{figure}[htbp]
\includegraphics[scale=0.35]{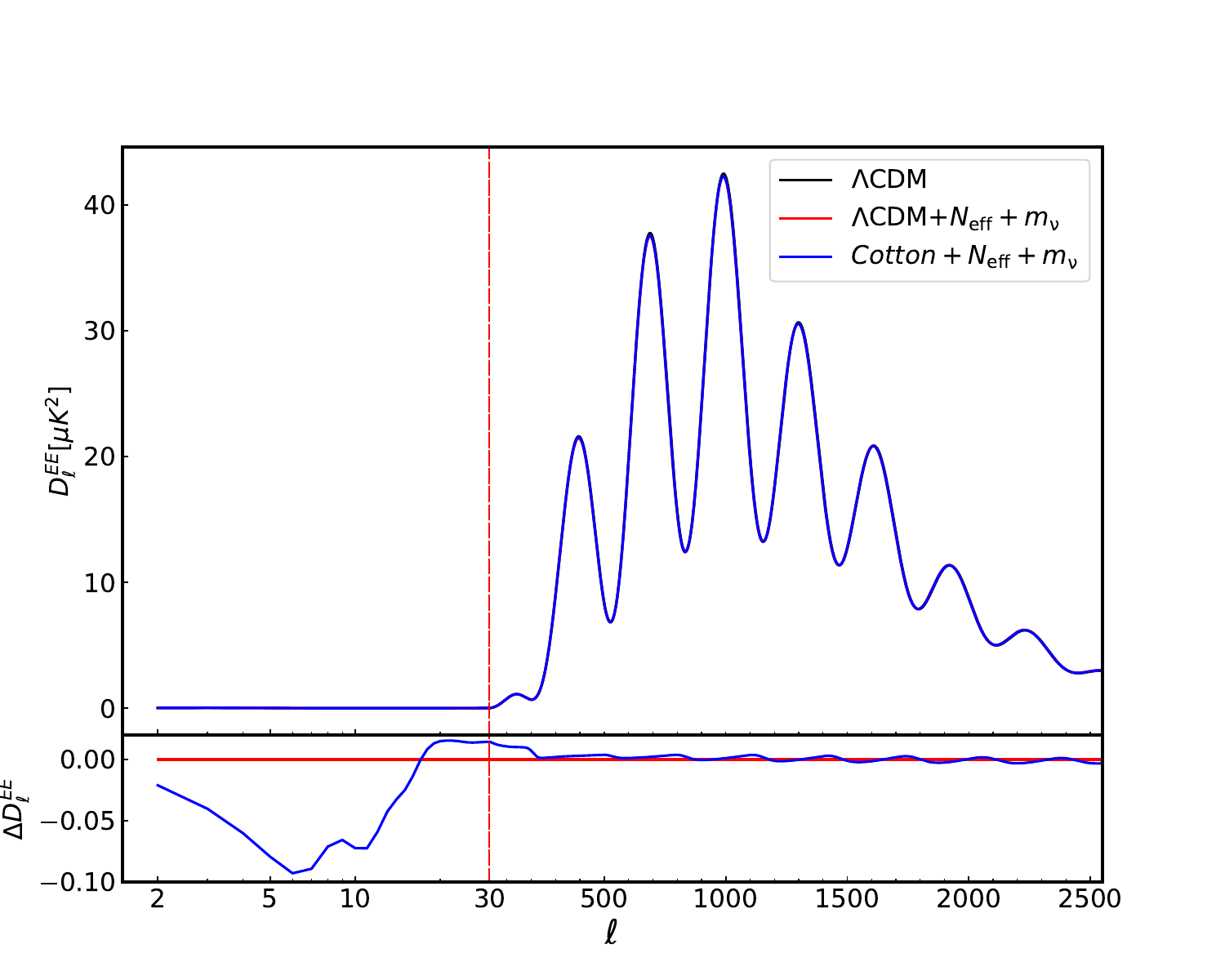}
\caption{The CMB $E$-mode temperature power spectra of $\Lambda$CDM, $\Lambda$CDM extension and Cotton gravity extension with their best-fit parameters. The difference between the latter two is shown in the lower panel.}
\label{fig:Cotton+neutrino_c_ee}
\end{figure}

The results are presented in Table \ref{tab:table2}. We shall compare Cotton gravity with $N_{\mathrm{eff}}$ and $m_{\mathrm{\nu}}$ added, as well as $\Lambda$CDM model with the same two parameters added. We will refer to these as ``Cotton gravity extension'' and ``$\Lambda$CDM extension'' respectively below. It can be observed that the values of neutrino effective parameters greatly influence the Cotton gravity parameters, as we expected. Specifically, the Planck 2018 data prefers the effective number of relativistic species to be smaller than 3 and the neutrino mass of less than 0.06eV. This explains why the Planck data favors the aforementioned non-zero value of $\beta$. To validate this argument rigorously, we restrict two parameters, $N_{\mathrm{eff}}$ and $m_{\mathrm{\nu}}$ to be their best-fit value in Table \ref{tab:table2} and perform MCMC analysis again. The result is shown in Table \ref{tab:table3}, which is similar to the result in Table \ref{tab:table2}. This provides assurance to our claim. After incorporating the neutrino model, the best-fit value of $\beta$ is now constrained to be

\begin{table}[htbp]
\begin{ruledtabular}
\begin{tabular}{c|c}
\textrm{}&
\textrm{Cotton gravity constrained $N_{\mathrm{eff}}$ and $m_{\mathrm{\nu}}$}\\
\colrule
$\beta$ & -0.0022$\pm$0.0018\\
$\Omega_{\mathrm{b}}h^2$ & 0.02249$\pm$0.00015 \\
$\Omega_{\mathrm{c}}h^2$ & 0.1196$\pm$0.0012 \\
$100\theta_{*}$&1.04177$\pm$0.00030\\
$\tau$&0.0495$\pm$0.0075\\
$\ln(10^{10}A_{\mathrm{s}})$&3.036$\pm$0.015\\
$n_{\mathrm{s}}$&0.9676$\pm$0.0043\\
$\chi^2$&1430
\end{tabular}
\end{ruledtabular}
\caption{1-$\sigma$ confidence interval of the parameters in the Cotton gravity constrained $N_{\mathrm{eff}}$ and $m_{\mathrm{\nu}}$}
\label{tab:table3}
\end{table}

\begin{equation}
    \beta=-0.0002^{+0.0017}_{-0.0020},
\end{equation} 
in 1-$\sigma$ confidence interval. Now we see that 1-$\sigma$ confidence interval of $\beta$ is strictly constrained around the neighborhood of GR value, $\beta=0$.

Fig.~\ref{fig:Cotton+neutrino_c_tt} displays the power spectrum of CMB temperature for both Cotton gravity and $\Lambda$CDM extensions, along with their respective best-fit parameters. The small magnitude of the difference, less than two percent, is evident in Fig.~\ref{fig:Cotton+neutrino_c_tt}. Cotton gravity primarily influences the spectrum in the low-$\ell$ region and have minimal impact in the high-$\ell$ region. It is worth noting that the acoustic peak positions closely align with those of the $\Lambda$CDM model.
Additionally, we show the $E$-mode power spectrum of CMB in Fig.~\ref{fig:Cotton+neutrino_c_ee}. The conclusion is basically the same as that in temperature power spectrum.

\begin{figure*}[htb]
\includegraphics[scale=0.3]{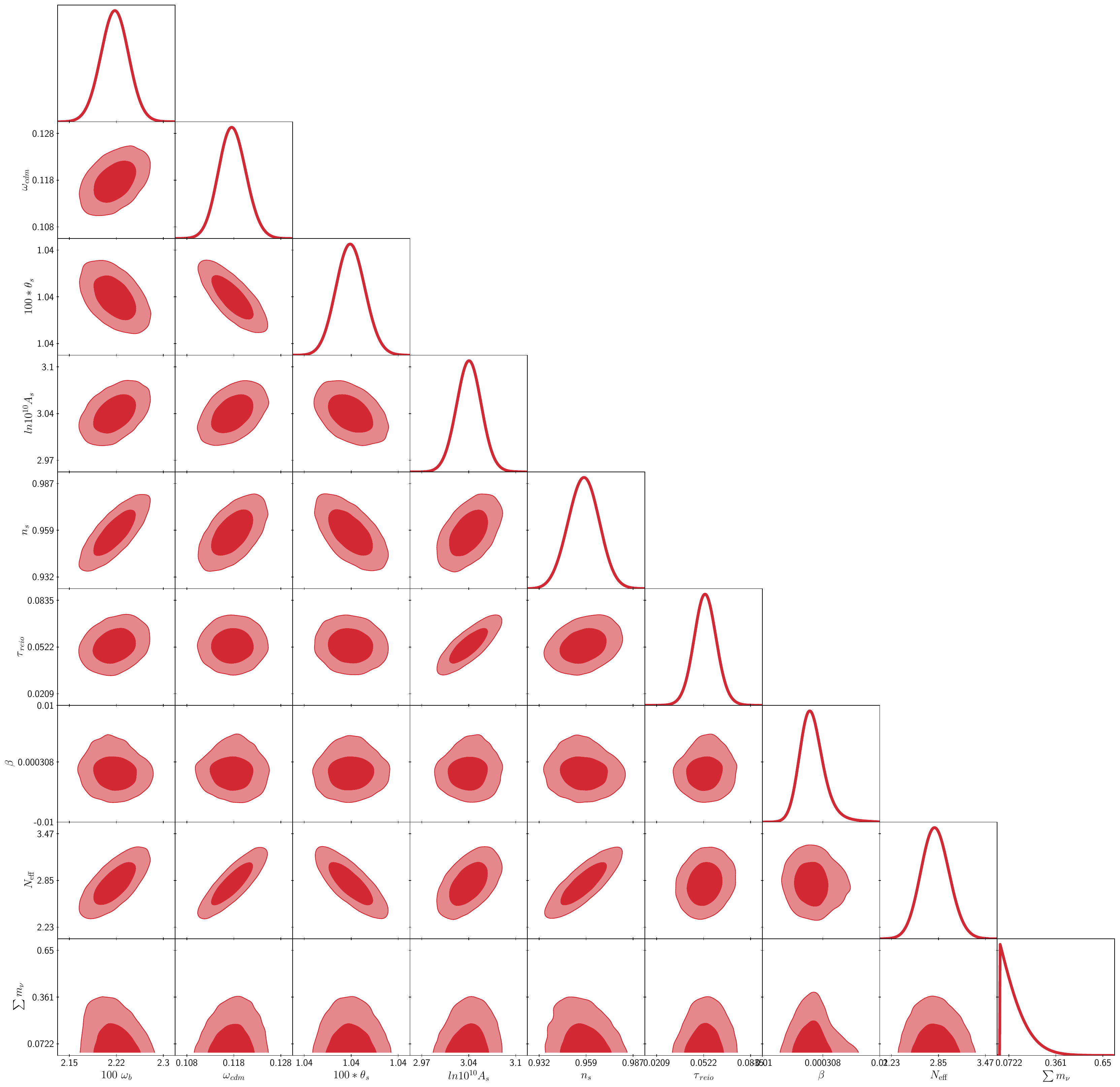}
\caption{Constraints on parameters of Cotton gravity with neutrino from the Planck 2018 data. Contours corresponding to 68\% and 95\% confidence levels.}
\label{fig:mcmc1}
\end{figure*}

The marginalized constraint contours for nine parameters, comprising of the standard six $\Lambda$CDM model parameters, two neutrino parameters and Cotton gravity parameter $\beta$, are presented in Fig.~\ref{fig:mcmc1}. The analysis of Fig.~\ref{fig:mcmc1} reveals that there is nearly no correlation between $\beta$ and $\theta_*$, as the perturbation theory of Cotton gravity has minimal impact on the comoving sound horizon. Additionally, the constraints imposed by the CMB amplitude result in only weak correlation between $\beta$ and the other parameters, especially $N_{\mathrm{eff}}$ and $m_{\mathrm{\nu}}$. The reason is that although Cotton gravity parameter $\beta$ and the anisotropic stress of the neutrino both influence the CMB, as the scale k becomes small, $\beta$ is invariant but the anisotropic stress decreases, approximately like $1/k^2$. At small scale they are totally different. This is why $N_{\mathrm{eff}}$ and $m_{\mathrm{\nu}}$ do not significantly influence $\beta$.

As mentioned before, the growth of matter in the late universe will be influenced by the value of Cotton gravity parameter. Specifically, we find that a tiny change of its value will cause a large change of matter power spectrum in the late universe. Additionally, the constraint from CMB data on Cotton gravity in high-$\ell$ region is mainly due to lensing, which is also affected by large scale structures. The data from large scale structures, which provides valuable information on the matter density fluctuations and CMB lensing, can potentially impose more stringent constraints on Cotton gravity \cite{Garcia-Arroyo:2020iou}. Therefore, it is worth combining the Planck and SDSS data \cite{SDSS:2008tqn} to constrain Cotton gravity. 

\begin{table}[htbp]
\begin{ruledtabular}
\begin{tabular}{c|c|c}
\textrm{}&
\textrm{Cotton gravity extension}& 
\textrm{$\Lambda$CDM + $N_{\mathrm{eff}}$ + $m_{\mathrm{\nu}}$}\\
\colrule
$\beta$ & -0.00008$\pm$0.00092 &\\
$\Omega_{\mathrm{b}}h^2$ & 0.02210$\pm$0.00022 &0.02213$\pm$0.00022\\
$\Omega_{\mathrm{c}}h^2$ & 0.1178$\pm$0.0029 &0.1160$\pm$0.0028\\
$100\theta_{*}$&1.04211$\pm$0.00052& 1.04247$\pm$0.00052\\
$\tau$&0.0532$\pm$0.0076&0.0543$\pm$0.0074\\
$\ln(10^{10}A_{\mathrm{s}})$&3.036$\pm$0.018&3.031$\pm$0.017\\
$n_{\mathrm{s}}$&0.9585$\pm$0.0085& 0.9598$\pm$0.0084\\
$\sum m_{\nu}$ & $\sum m_{\nu}\text{\textless}$0.22 (95$\%$) & $\sum m_{\nu}\text{\textless}$0.21 (95$\%$)\\
$N_{\mathrm{eff}}$ & 2.81$\pm$0.35 (95$\%$) & 2.77$\pm$0.36 (95$\%$)\\
$\chi^2$& 1411 & 1411
\end{tabular}
\end{ruledtabular}
\caption{1-$\sigma$ confidence interval of the parameters in Cotton gravity plus neutrino and in $\Lambda$CDM model (combined Planck and SDSS data).}
\label{tab:table3}
\end{table}

The results are shown in Table \ref{tab:table3}. The Cotton gravity parameter is now even more tightly constrained. Its best-fit value becomes 
\begin{equation}
     \beta=-0.00008^{+0.00080}_{-0.00104},
\end{equation}
within 1-$\sigma$ confidence interval.
The neutrino mass is also better constrained due to its effect on the growth of matter power spectrum. Other parameters are hardly affected.
\begin{figure}[htbp]
\includegraphics[scale=0.30]{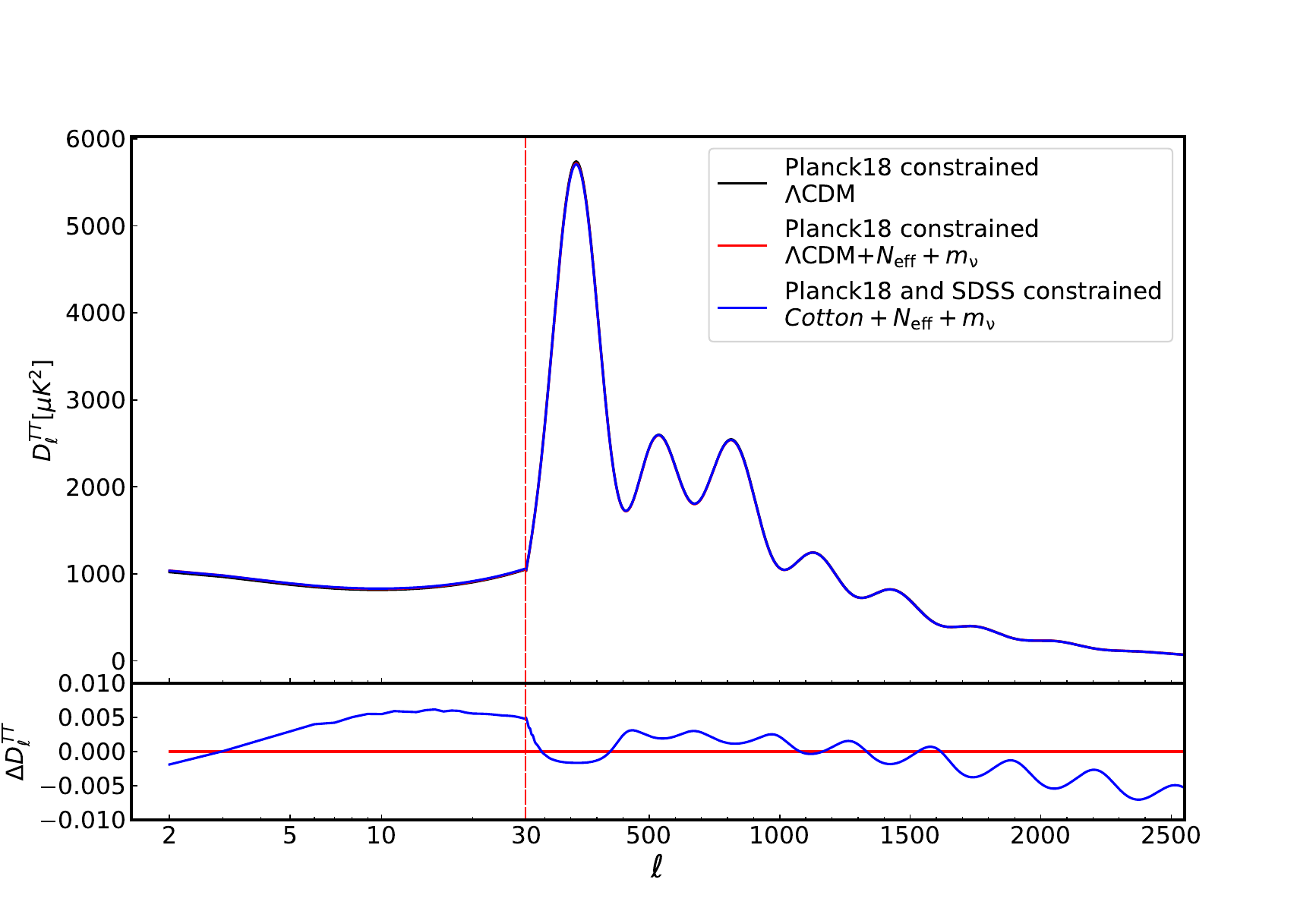}
\caption{The CMB temperature power spectrum of $\Lambda$CDM, $\Lambda$CDM extension and Cotton gravity extension with their best-fit parameters. The difference between the latter two is shown in the lower panel.}
\label{fig:Cotton+sdss+neutrino_c_tt+delta}
\end{figure}
\begin{figure}[htbp]
\includegraphics[scale=0.30]{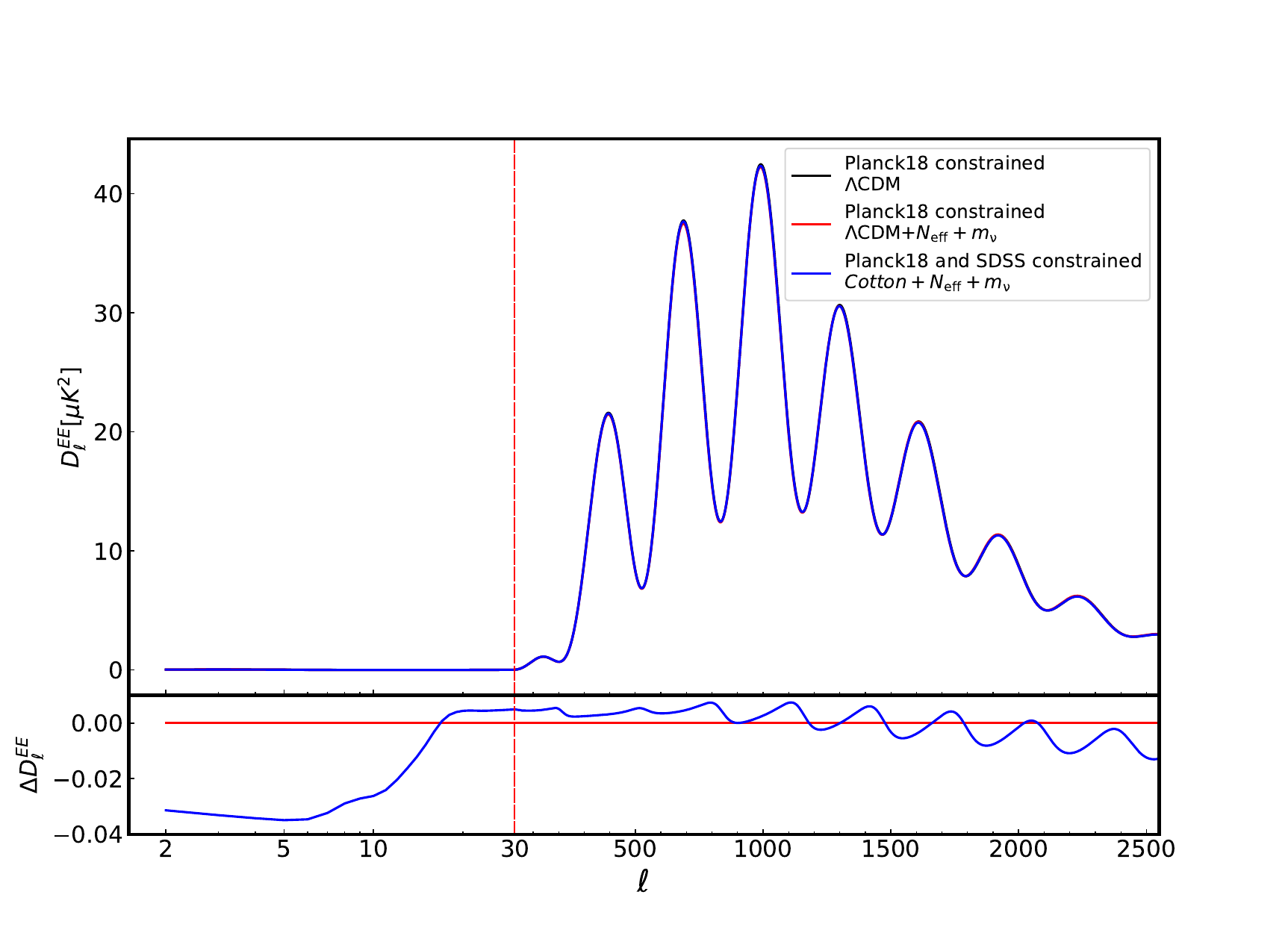}
\caption{The CMB $E$-mode polarization power spectrum of $\Lambda$CDM, $\Lambda$CDM extension and Cotton gravity extension with their best-fit parameters. The difference between the latter two is shown in the lower panel.}
\label{fig:Cotton+sdss+neutrino_c_ee+delta}
\end{figure}
\begin{figure*}[htbp]
\includegraphics[scale=0.3]{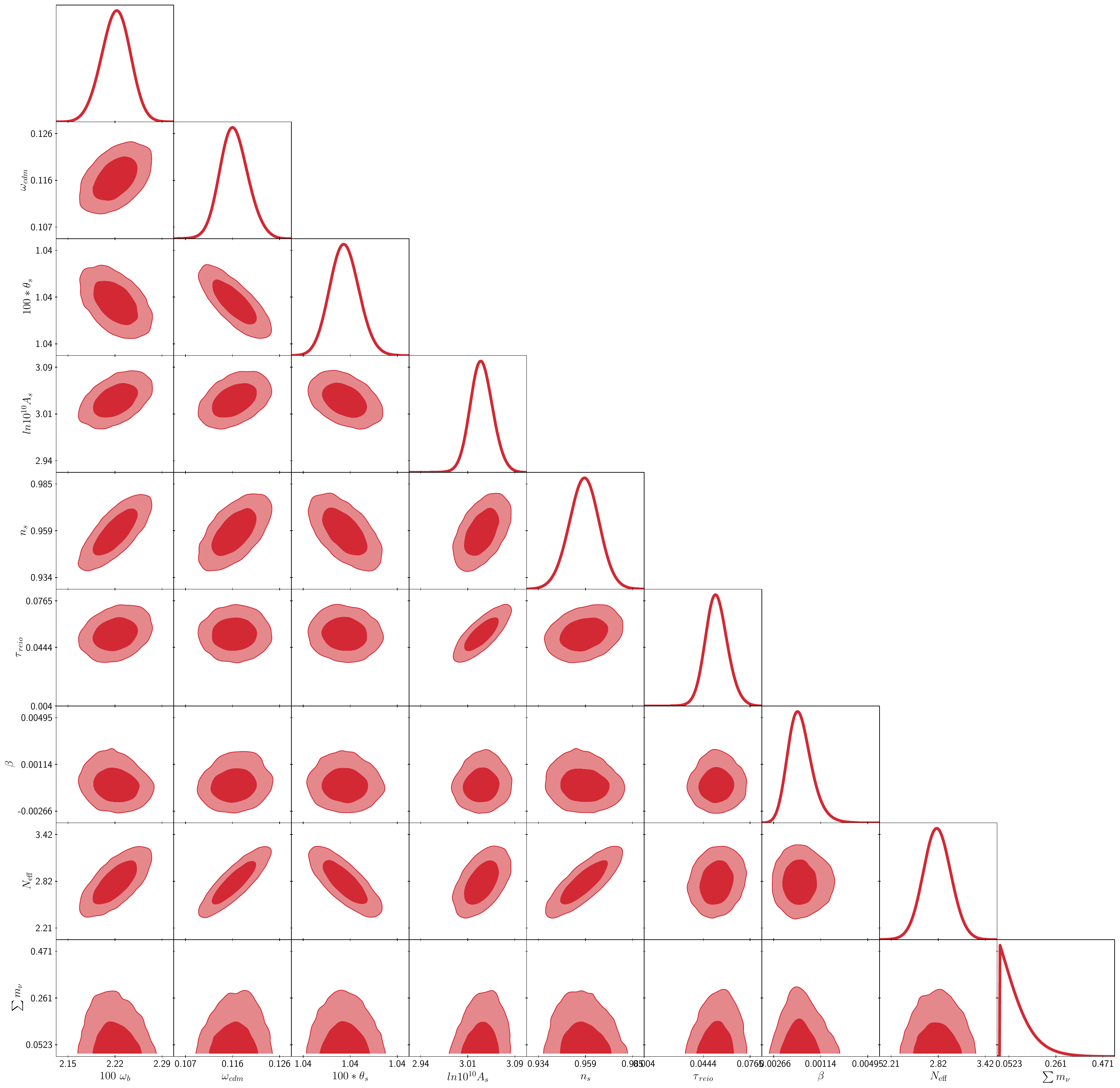}
\caption{Constraints on parameters of Cotton gravity extension model from the Planck 2018 and SDSS data. Contours corresponding to 68\% and 95\% confidence levels.}
\label{fig:mcmc2}
\end{figure*}

We also display the temperature power spectrum of CMB in Fig.~\ref{fig:Cotton+sdss+neutrino_c_tt+delta} and $E$-mode power spectrum in Fig.~\ref{fig:Cotton+sdss+neutrino_c_ee+delta}. Their deviation from $\Lambda$CDM extension model becomes smaller than the case using only Planck's data, which is less than four percent. Finally, the marginalized constraint contours are presented in Fig.~\ref{fig:mcmc2}. Other conclusions remain unchanged, so we will not repeat here.

\section{CONCLUSION: No EVIDENCE FOR COTTON GRAVITY\label{conclusion}}
In this work we have investigated the cosmological perturbation in Cotton gravity theory \cite{Harada:2021bte}. This modification in gravity is intriguing as it extends the principles of GR and encompasses all solutions associated with GR (in particular, it can also incorporate a cosmological constant). However, the theory goes beyond GR in that it may provide an explanation for the galaxy rotation curve without invoking the presence of dark matter \cite{Harada:2022edl}.

However, galaxy rotation curve is not the only aspect one should test any gravity theory that purportedly can replace dark matter. 
In this study, due to the lack of uniqueness evolution in Cotton gravity even at the background level, we have made the "minimal assumption" that the background is flat FLRW spacetime as in GR. In other words, our approach is phenomenological -- we seek to study the possible effects of Cotton gravity beyond the background level after this assumption is made. our focus is to analyze Cotton gravity's influence on CMB power spectrum and large scale structures. We have discovered that the effect of Cotton gravity can be treated as an anisotropic fluid, which in turn can be described by an arbitrary function, $\beta(k)$, at the linear perturbation level. Consequently, the gravitational potentials will evolve differently compared to the $\Lambda$CDM model (even if the background is assumed to be the same), thereby impacting the CMB and large scale structures. For simplicity, in this article, we only consider the case that the parameter $\beta$ remains constant.

To study the evolution of perturbations in Cotton gravity, we have made modifications to the Einstein-Boltzmann solver, MGCLASS II \cite{Sakr:2021ylx}. Additionally, we conducted an MCMC simulation to constrain the six cosmological parameters and Cotton gravity parameter $\beta$, and determine their best-fit values by using Planck data. The best-fit value of $\beta$ from CMB alone is $0.110^{+0.011}_{-0.010}$, which seemingly deviates from zero at a significance level of at least 3$\sigma$. However, the disparities in matter power spectra and CMB lensing-potential power spectra lead us to believe $\beta$ cannot deviate too much from zero. Note that since the effect of Cotton gravity is equivalent to an anisotropic fluid, other physical processes which influence the anisotropic stress will affect the best-fit value of $\beta$. 

If two additional parameters, namely $N_{\mathrm{eff}}$ and the neutrino mass $m_{\nu}$, is taken into the MCMC analysis, the best-fit value of $\beta$ reduces to $-0.0002^{+0.0017}_{-0.0020}$ in 1-$\sigma$ confidence interval. The discrepancy of the two different best-fits is due to the fact that the Planck data favors an effective number of  $N_{\mathrm{eff}}$ less than three. If we combine the SDSS data from large scale structures, the constraint will become stricter. The best-fit value of $\beta$ becomes $-0.00008^{+0.00080}_{-0.00104}$, which is even closer to zero.

Therefore, to conclude, based on our current findings, at least from a cosmological point of view, Cotton gravity model with a constant parameter $\beta$ does \emph{not} exhibit any deviation from GR. In other words, there is no evidence for such a modification of gravity.
The possibility remains that the Cotton gravity parameter $\beta$ in Eq. \eqref{eq:111} is not a constant, in which case the analysis would be more challenging but perhaps might lead to interesting physics. It would therefore be interesting to constrain such a time-varying or energy-scale varying $\beta$ in future studies. 

\begin{acknowledgments}
We are grateful to Yi-Fu Cai, Elisa Ferreira, Qingqing Wang, and Geyu Mo for helpful discussions. This work is supported in part by the National Key R\&D Program of China (2021YFC2203100), CAS Young Interdisciplinary Innovation Team (JCTD-2022-20), NSFC (12261131497, 12003029), 111 Project for ``Observational and Theoretical Research on Dark Matter and Dark Energy" (B23042), by Fundamental Research Funds for Central Universities, by CSC Innovation Talent Funds, by USTC Fellowships for International Cooperation, and by USTC Research Funds of the Double First-Class Initiative. Kavli IPMU is supported by World Premier International Research Center Initiative (WPI), MEXT, Japan. We acknowledge the use of computing facilities of astronomy department, as well as the clusters LINDA \& JUDY of the particle cosmology group at USTC. 
\end{acknowledgments}

\appendix
\section{Code Detail\label{appendix}}
The code we used to calculate the CMB power spectrum is based on MGCLASS II \cite{Sakr:2021ylx}. MGCLASS II makes phenomenological paramerization of Poisson and anisotropic stress equations to describe the difference between modified gravity and GR at the cosmological perturbation level. Since we assume the background evolution of Cotton gravity is the same as flat FLRW background, we did not change the code of MGCLASS II in the background part.
We only imitate MGCLASS II to change the perturbation equation. To be Special, we rewrite the code about cosmology perturbation function in $\texttt{Class/source/perturbation.c}$ document and put in the parameter $\beta$ in code.

\bibliography{apssamp}

\providecommand{\noopsort}[1]{}\providecommand{\singleletter}[1]{#1}%
\begin{thebibliography}{48}%
\makeatletter
\providecommand \@ifxundefined [1]{%
 \@ifx{#1\undefined}
}%
\providecommand \@ifnum [1]{%
 \ifnum #1\expandafter \@firstoftwo
 \else \expandafter \@secondoftwo
 \fi
}%
\providecommand \@ifx [1]{%
 \ifx #1\expandafter \@firstoftwo
 \else \expandafter \@secondoftwo
 \fi
}%
\providecommand \natexlab [1]{#1}%
\providecommand \enquote  [1]{``#1''}%
\providecommand \bibnamefont  [1]{#1}%
\providecommand \bibfnamefont [1]{#1}%
\providecommand \citenamefont [1]{#1}%
\providecommand \href@noop [0]{\@secondoftwo}%
\providecommand \href [0]{\begingroup \@sanitize@url \@href}%
\providecommand \@href[1]{\@@startlink{#1}\@@href}%
\providecommand \@@href[1]{\endgroup#1\@@endlink}%
\providecommand \@sanitize@url [0]{\catcode `\\12\catcode `\$12\catcode `\&12\catcode `\#12\catcode `\^12\catcode `\_12\catcode `\%12\relax}%
\providecommand \@@startlink[1]{}%
\providecommand \@@endlink[0]{}%
\providecommand \url  [0]{\begingroup\@sanitize@url \@url }%
\providecommand \@url [1]{\endgroup\@href {#1}{\urlprefix }}%
\providecommand \urlprefix  [0]{URL }%
\providecommand \Eprint [0]{\href }%
\providecommand \doibase [0]{https://doi.org/}%
\providecommand \selectlanguage [0]{\@gobble}%
\providecommand \bibinfo  [0]{\@secondoftwo}%
\providecommand \bibfield  [0]{\@secondoftwo}%
\providecommand \translation [1]{[#1]}%
\providecommand \BibitemOpen [0]{}%
\providecommand \bibitemStop [0]{}%
\providecommand \bibitemNoStop [0]{.\EOS\space}%
\providecommand \EOS [0]{\spacefactor3000\relax}%
\providecommand \BibitemShut  [1]{\csname bibitem#1\endcsname}%
\let\auto@bib@innerbib\@empty
\bibitem [{\citenamefont {Will}(2014)}]{Will:2014kxa}%
  \BibitemOpen
  \bibfield  {author} {\bibinfo {author} {\bibfnamefont {C.~M.}\ \bibnamefont {Will}},\ }\bibfield  {title} {\bibinfo {title} {{The confrontation between general relativity and experiment}},\ }\href {https://doi.org/10.12942/lrr-2014-4} {\bibfield  {journal} {\bibinfo  {journal} {Living Rev. Rel.}\ }\textbf {\bibinfo {volume} {17}},\ \bibinfo {pages} {4} (\bibinfo {year} {2014})},\ \Eprint {https://arxiv.org/abs/1403.7377} {arXiv:1403.7377 [gr-qc]} \BibitemShut {NoStop}%
\bibitem [{\citenamefont {Ishak}(2019)}]{Ishak:2018his}%
  \BibitemOpen
  \bibfield  {author} {\bibinfo {author} {\bibfnamefont {M.}~\bibnamefont {Ishak}},\ }\bibfield  {title} {\bibinfo {title} {{Testing general relativity in cosmology}},\ }\href {https://doi.org/10.1007/s41114-018-0017-4} {\bibfield  {journal} {\bibinfo  {journal} {Living Rev. Rel.}\ }\textbf {\bibinfo {volume} {22}},\ \bibinfo {pages} {1} (\bibinfo {year} {2019})},\ \Eprint {https://arxiv.org/abs/1806.10122} {arXiv:1806.10122 [astro-ph.CO]} \BibitemShut {NoStop}%
\bibitem [{\citenamefont {Berti}\ \emph {et~al.}(2015)\citenamefont {Berti} \emph {et~al.}}]{Berti:2015itd}%
  \BibitemOpen
  \bibfield  {author} {\bibinfo {author} {\bibfnamefont {E.}~\bibnamefont {Berti}} \emph {et~al.},\ }\bibfield  {title} {\bibinfo {title} {{Testing general relativity with present and future astrophysical observations}},\ }\href {https://doi.org/10.1088/0264-9381/32/24/243001} {\bibfield  {journal} {\bibinfo  {journal} {Class. Quant. Grav.}\ }\textbf {\bibinfo {volume} {32}},\ \bibinfo {pages} {243001} (\bibinfo {year} {2015})},\ \Eprint {https://arxiv.org/abs/1501.07274} {arXiv:1501.07274 [gr-qc]} \BibitemShut {NoStop}%
\bibitem [{\citenamefont {Hinshaw}\ \emph {et~al.}(2003)\citenamefont {Hinshaw} \emph {et~al.}}]{WMAP:2003zzr}%
  \BibitemOpen
  \bibfield  {author} {\bibinfo {author} {\bibfnamefont {G.}~\bibnamefont {Hinshaw}} \emph {et~al.} (\bibinfo {collaboration} {WMAP}),\ }\bibfield  {title} {\bibinfo {title} {{First year Wilkinson Microwave Anisotropy Probe (WMAP) observations: the Angular power spectrum}},\ }\href {https://doi.org/10.1086/377225} {\bibfield  {journal} {\bibinfo  {journal} {Astrophys. J. Suppl.}\ }\textbf {\bibinfo {volume} {148}},\ \bibinfo {pages} {135} (\bibinfo {year} {2003})},\ \Eprint {https://arxiv.org/abs/astro-ph/0302217} {arXiv:astro-ph/0302217} \BibitemShut {NoStop}%
\bibitem [{\citenamefont {Aghanim}\ \emph {et~al.}(2020{\natexlab{a}})\citenamefont {Aghanim} \emph {et~al.}}]{Planck:2018vyg}%
  \BibitemOpen
  \bibfield  {author} {\bibinfo {author} {\bibfnamefont {N.}~\bibnamefont {Aghanim}} \emph {et~al.} (\bibinfo {collaboration} {Planck}),\ }\bibfield  {title} {\bibinfo {title} {{Planck 2018 results. VI. Cosmological parameters}},\ }\href {https://doi.org/10.1051/0004-6361/201833910} {\bibfield  {journal} {\bibinfo  {journal} {Astron. Astrophys.}\ }\textbf {\bibinfo {volume} {641}},\ \bibinfo {pages} {A6} (\bibinfo {year} {2020}{\natexlab{a}})},\ \bibinfo {note} {[Erratum: Astron.Astrophys. 652, C4 (2021)]},\ \Eprint {https://arxiv.org/abs/1807.06209} {arXiv:1807.06209 [astro-ph.CO]} \BibitemShut {NoStop}%
\bibitem [{\citenamefont {Rubin}\ \emph {et~al.}(1980)\citenamefont {Rubin}, \citenamefont {Thonnard},\ and\ \citenamefont {Ford}}]{Rubin:1980zd}%
  \BibitemOpen
  \bibfield  {author} {\bibinfo {author} {\bibfnamefont {V.~C.}\ \bibnamefont {Rubin}}, \bibinfo {author} {\bibfnamefont {N.}~\bibnamefont {Thonnard}},\ and\ \bibinfo {author} {\bibfnamefont {W.~K.}\ \bibnamefont {Ford}, \bibfnamefont {Jr.}},\ }\bibfield  {title} {\bibinfo {title} {{Rotational properties of 21 SC galaxies with a large range of luminosities and radii, from NGC 4605 /R = 4kpc/ to UGC 2885 /R = 122 kpc/}},\ }\href {https://doi.org/10.1086/158003} {\bibfield  {journal} {\bibinfo  {journal} {Astrophys. J.}\ }\textbf {\bibinfo {volume} {238}},\ \bibinfo {pages} {471} (\bibinfo {year} {1980})}\BibitemShut {NoStop}%
\bibitem [{\citenamefont {Lelli}\ \emph {et~al.}(2016)\citenamefont {Lelli}, \citenamefont {McGaugh},\ and\ \citenamefont {Schombert}}]{Lelli:2016zqa}%
  \BibitemOpen
  \bibfield  {author} {\bibinfo {author} {\bibfnamefont {F.}~\bibnamefont {Lelli}}, \bibinfo {author} {\bibfnamefont {S.~S.}\ \bibnamefont {McGaugh}},\ and\ \bibinfo {author} {\bibfnamefont {J.~M.}\ \bibnamefont {Schombert}},\ }\bibfield  {title} {\bibinfo {title} {{SPARC: Mass models for 175 disk galaxies with spitzer photometry and accurate rotation curves}},\ }\href {https://doi.org/10.3847/0004-6256/152/6/157} {\bibfield  {journal} {\bibinfo  {journal} {Astron. J.}\ }\textbf {\bibinfo {volume} {152}},\ \bibinfo {pages} {157} (\bibinfo {year} {2016})},\ \Eprint {https://arxiv.org/abs/1606.09251} {arXiv:1606.09251 [astro-ph.GA]} \BibitemShut {NoStop}%
\bibitem [{\citenamefont {Riess}\ \emph {et~al.}(1998)\citenamefont {Riess} \emph {et~al.}}]{SupernovaSearchTeam:1998fmf}%
  \BibitemOpen
  \bibfield  {author} {\bibinfo {author} {\bibfnamefont {A.~G.}\ \bibnamefont {Riess}} \emph {et~al.} (\bibinfo {collaboration} {Supernova Search Team}),\ }\bibfield  {title} {\bibinfo {title} {{Observational evidence from supernovae for an accelerating universe and a cosmological constant}},\ }\href {https://doi.org/10.1086/300499} {\bibfield  {journal} {\bibinfo  {journal} {Astron. J.}\ }\textbf {\bibinfo {volume} {116}},\ \bibinfo {pages} {1009} (\bibinfo {year} {1998})},\ \Eprint {https://arxiv.org/abs/astro-ph/9805201} {arXiv:astro-ph/9805201} \BibitemShut {NoStop}%
\bibitem [{\citenamefont {Riess}\ \emph {et~al.}(2022)\citenamefont {Riess} \emph {et~al.}}]{Riess:2021jrx}%
  \BibitemOpen
  \bibfield  {author} {\bibinfo {author} {\bibfnamefont {A.~G.}\ \bibnamefont {Riess}} \emph {et~al.},\ }\bibfield  {title} {\bibinfo {title} {{A comprehensive measurement of the local value of the Hubble constant with 1 km/s/Mpc uncertainty from the Hubble Space Telescope and the SH0ES Team}},\ }\href {https://doi.org/10.3847/2041-8213/ac5c5b} {\bibfield  {journal} {\bibinfo  {journal} {Astrophys. J. Lett.}\ }\textbf {\bibinfo {volume} {934}},\ \bibinfo {pages} {L7} (\bibinfo {year} {2022})},\ \Eprint {https://arxiv.org/abs/2112.04510} {arXiv:2112.04510 [astro-ph.CO]} \BibitemShut {NoStop}%
\bibitem [{\citenamefont {Adame}\ \emph {et~al.}(2024)\citenamefont {Adame} \emph {et~al.}}]{DESI:2024mwx}%
  \BibitemOpen
  \bibfield  {author} {\bibinfo {author} {\bibfnamefont {A.~G.}\ \bibnamefont {Adame}} \emph {et~al.} (\bibinfo {collaboration} {DESI}),\ }\bibfield  {title} {\bibinfo {title} {{DESI 2024 VI: Cosmological constraints from the measurements of Baryon Acoustic Oscillations}},\ }\href@noop {} {\  (\bibinfo {year} {2024})},\ \Eprint {https://arxiv.org/abs/2404.03002} {arXiv:2404.03002 [astro-ph.CO]} \BibitemShut {NoStop}%
\bibitem [{\citenamefont {Percival}\ \emph {et~al.}(2010)\citenamefont {Percival} \emph {et~al.}}]{SDSS:2009ocz}%
  \BibitemOpen
  \bibfield  {author} {\bibinfo {author} {\bibfnamefont {W.~J.}\ \bibnamefont {Percival}} \emph {et~al.} (\bibinfo {collaboration} {SDSS}),\ }\bibfield  {title} {\bibinfo {title} {{Baryon Acoustic Oscillations in the Sloan Digital Sky Survey Data Release 7 Galaxy Sample}},\ }\href {https://doi.org/10.1111/j.1365-2966.2009.15812.x} {\bibfield  {journal} {\bibinfo  {journal} {Mon. Not. Roy. Astron. Soc.}\ }\textbf {\bibinfo {volume} {401}},\ \bibinfo {pages} {2148} (\bibinfo {year} {2010})},\ \Eprint {https://arxiv.org/abs/0907.1660} {arXiv:0907.1660 [astro-ph.CO]} \BibitemShut {NoStop}%
\bibitem [{\citenamefont {Di~Valentino}\ \emph {et~al.}(2021)\citenamefont {Di~Valentino}, \citenamefont {Mena}, \citenamefont {Pan}, \citenamefont {Visinelli}, \citenamefont {Yang}, \citenamefont {Melchiorri}, \citenamefont {Mota}, \citenamefont {Riess},\ and\ \citenamefont {Silk}}]{DiValentino:2021izs}%
  \BibitemOpen
  \bibfield  {author} {\bibinfo {author} {\bibfnamefont {E.}~\bibnamefont {Di~Valentino}}, \bibinfo {author} {\bibfnamefont {O.}~\bibnamefont {Mena}}, \bibinfo {author} {\bibfnamefont {S.}~\bibnamefont {Pan}}, \bibinfo {author} {\bibfnamefont {L.}~\bibnamefont {Visinelli}}, \bibinfo {author} {\bibfnamefont {W.}~\bibnamefont {Yang}}, \bibinfo {author} {\bibfnamefont {A.}~\bibnamefont {Melchiorri}}, \bibinfo {author} {\bibfnamefont {D.~F.}\ \bibnamefont {Mota}}, \bibinfo {author} {\bibfnamefont {A.~G.}\ \bibnamefont {Riess}},\ and\ \bibinfo {author} {\bibfnamefont {J.}~\bibnamefont {Silk}},\ }\bibfield  {title} {\bibinfo {title} {{In the realm of the Hubble tension\textemdash{}a review of solutions}},\ }\href {https://doi.org/10.1088/1361-6382/ac086d} {\bibfield  {journal} {\bibinfo  {journal} {Class. Quant. Grav.}\ }\textbf {\bibinfo {volume} {38}},\ \bibinfo {pages} {153001} (\bibinfo {year} {2021})},\ \Eprint {https://arxiv.org/abs/2103.01183} {arXiv:2103.01183 [astro-ph.CO]} \BibitemShut {NoStop}%
\bibitem [{\citenamefont {Horndeski}(1974)}]{Horndeski:1974wa}%
  \BibitemOpen
  \bibfield  {author} {\bibinfo {author} {\bibfnamefont {G.~W.}\ \bibnamefont {Horndeski}},\ }\bibfield  {title} {\bibinfo {title} {{Second-order scalar-tensor field equations in a four-dimensional space}},\ }\href {https://doi.org/10.1007/BF01807638} {\bibfield  {journal} {\bibinfo  {journal} {Int. J. Theor. Phys.}\ }\textbf {\bibinfo {volume} {10}},\ \bibinfo {pages} {363} (\bibinfo {year} {1974})}\BibitemShut {NoStop}%
\bibitem [{\citenamefont {Khoury}\ and\ \citenamefont {Weltman}(2004)}]{Khoury:2003rn}%
  \BibitemOpen
  \bibfield  {author} {\bibinfo {author} {\bibfnamefont {J.}~\bibnamefont {Khoury}}\ and\ \bibinfo {author} {\bibfnamefont {A.}~\bibnamefont {Weltman}},\ }\bibfield  {title} {\bibinfo {title} {{Chameleon cosmology}},\ }\href {https://doi.org/10.1103/PhysRevD.69.044026} {\bibfield  {journal} {\bibinfo  {journal} {Phys. Rev. D}\ }\textbf {\bibinfo {volume} {69}},\ \bibinfo {pages} {044026} (\bibinfo {year} {2004})},\ \Eprint {https://arxiv.org/abs/astro-ph/0309411} {arXiv:astro-ph/0309411} \BibitemShut {NoStop}%
\bibitem [{\citenamefont {Langlois}\ and\ \citenamefont {Noui}(2016)}]{Langlois:2015cwa}%
  \BibitemOpen
  \bibfield  {author} {\bibinfo {author} {\bibfnamefont {D.}~\bibnamefont {Langlois}}\ and\ \bibinfo {author} {\bibfnamefont {K.}~\bibnamefont {Noui}},\ }\bibfield  {title} {\bibinfo {title} {{Degenerate higher derivative theories beyond Horndeski: evading the Ostrogradski instability}},\ }\href {https://doi.org/10.1088/1475-7516/2016/02/034} {\bibfield  {journal} {\bibinfo  {journal} {JCAP}\ }\textbf {\bibinfo {volume} {02}},\ \bibinfo {pages} {034}},\ \Eprint {https://arxiv.org/abs/1510.06930} {arXiv:1510.06930 [gr-qc]} \BibitemShut {NoStop}%
\bibitem [{\citenamefont {Ben~Achour}\ \emph {et~al.}(2016)\citenamefont {Ben~Achour}, \citenamefont {Langlois},\ and\ \citenamefont {Noui}}]{BenAchour:2016cay}%
  \BibitemOpen
  \bibfield  {author} {\bibinfo {author} {\bibfnamefont {J.}~\bibnamefont {Ben~Achour}}, \bibinfo {author} {\bibfnamefont {D.}~\bibnamefont {Langlois}},\ and\ \bibinfo {author} {\bibfnamefont {K.}~\bibnamefont {Noui}},\ }\bibfield  {title} {\bibinfo {title} {{Degenerate higher order scalar-tensor theories beyond Horndeski and disformal transformations}},\ }\href {https://doi.org/10.1103/PhysRevD.93.124005} {\bibfield  {journal} {\bibinfo  {journal} {Phys. Rev. D}\ }\textbf {\bibinfo {volume} {93}},\ \bibinfo {pages} {124005} (\bibinfo {year} {2016})},\ \Eprint {https://arxiv.org/abs/1602.08398} {arXiv:1602.08398 [gr-qc]} \BibitemShut {NoStop}%
\bibitem [{\citenamefont {Cai}\ \emph {et~al.}(2016)\citenamefont {Cai}, \citenamefont {Capozziello}, \citenamefont {De~Laurentis},\ and\ \citenamefont {Saridakis}}]{Cai:2015emx}%
  \BibitemOpen
  \bibfield  {author} {\bibinfo {author} {\bibfnamefont {Y.-F.}\ \bibnamefont {Cai}}, \bibinfo {author} {\bibfnamefont {S.}~\bibnamefont {Capozziello}}, \bibinfo {author} {\bibfnamefont {M.}~\bibnamefont {De~Laurentis}},\ and\ \bibinfo {author} {\bibfnamefont {E.~N.}\ \bibnamefont {Saridakis}},\ }\bibfield  {title} {\bibinfo {title} {{f(T) teleparallel gravity and cosmology}},\ }\href {https://doi.org/10.1088/0034-4885/79/10/106901} {\bibfield  {journal} {\bibinfo  {journal} {Rept. Prog. Phys.}\ }\textbf {\bibinfo {volume} {79}},\ \bibinfo {pages} {106901} (\bibinfo {year} {2016})},\ \Eprint {https://arxiv.org/abs/1511.07586} {arXiv:1511.07586 [gr-qc]} \BibitemShut {NoStop}%
\bibitem [{\citenamefont {Krssak}\ \emph {et~al.}(2019)\citenamefont {Krssak}, \citenamefont {van~den Hoogen}, \citenamefont {Pereira}, \citenamefont {B\"ohmer},\ and\ \citenamefont {Coley}}]{Krssak:2018ywd}%
  \BibitemOpen
  \bibfield  {author} {\bibinfo {author} {\bibfnamefont {M.}~\bibnamefont {Krssak}}, \bibinfo {author} {\bibfnamefont {R.~J.}\ \bibnamefont {van~den Hoogen}}, \bibinfo {author} {\bibfnamefont {J.~G.}\ \bibnamefont {Pereira}}, \bibinfo {author} {\bibfnamefont {C.~G.}\ \bibnamefont {B\"ohmer}},\ and\ \bibinfo {author} {\bibfnamefont {A.~A.}\ \bibnamefont {Coley}},\ }\bibfield  {title} {\bibinfo {title} {{Teleparallel theories of gravity: illuminating a fully invariant approach}},\ }\href {https://doi.org/10.1088/1361-6382/ab2e1f} {\bibfield  {journal} {\bibinfo  {journal} {Class. Quant. Grav.}\ }\textbf {\bibinfo {volume} {36}},\ \bibinfo {pages} {183001} (\bibinfo {year} {2019})},\ \Eprint {https://arxiv.org/abs/1810.12932} {arXiv:1810.12932 [gr-qc]} \BibitemShut {NoStop}%
\bibitem [{\citenamefont {Cai}\ \emph {et~al.}(2011)\citenamefont {Cai}, \citenamefont {Chen}, \citenamefont {Dent}, \citenamefont {Dutta},\ and\ \citenamefont {Saridakis}}]{Cai:2011tc}%
  \BibitemOpen
  \bibfield  {author} {\bibinfo {author} {\bibfnamefont {Y.-F.}\ \bibnamefont {Cai}}, \bibinfo {author} {\bibfnamefont {S.-H.}\ \bibnamefont {Chen}}, \bibinfo {author} {\bibfnamefont {J.~B.}\ \bibnamefont {Dent}}, \bibinfo {author} {\bibfnamefont {S.}~\bibnamefont {Dutta}},\ and\ \bibinfo {author} {\bibfnamefont {E.~N.}\ \bibnamefont {Saridakis}},\ }\bibfield  {title} {\bibinfo {title} {{Matter bounce cosmology with the f(T) gravity}},\ }\href {https://doi.org/10.1088/0264-9381/28/21/215011} {\bibfield  {journal} {\bibinfo  {journal} {Class. Quant. Grav.}\ }\textbf {\bibinfo {volume} {28}},\ \bibinfo {pages} {215011} (\bibinfo {year} {2011})},\ \Eprint {https://arxiv.org/abs/1104.4349} {arXiv:1104.4349 [astro-ph.CO]} \BibitemShut {NoStop}%
\bibitem [{\citenamefont {Bengochea}\ and\ \citenamefont {Ferraro}(2009)}]{Bengochea:2008gz}%
  \BibitemOpen
  \bibfield  {author} {\bibinfo {author} {\bibfnamefont {G.~R.}\ \bibnamefont {Bengochea}}\ and\ \bibinfo {author} {\bibfnamefont {R.}~\bibnamefont {Ferraro}},\ }\bibfield  {title} {\bibinfo {title} {{Dark torsion as the cosmic speed-up}},\ }\href {https://doi.org/10.1103/PhysRevD.79.124019} {\bibfield  {journal} {\bibinfo  {journal} {Phys. Rev. D}\ }\textbf {\bibinfo {volume} {79}},\ \bibinfo {pages} {124019} (\bibinfo {year} {2009})},\ \Eprint {https://arxiv.org/abs/0812.1205} {arXiv:0812.1205 [astro-ph]} \BibitemShut {NoStop}%
\bibitem [{\citenamefont {Sotiriou}\ and\ \citenamefont {Faraoni}(2010)}]{Sotiriou:2008rp}%
  \BibitemOpen
  \bibfield  {author} {\bibinfo {author} {\bibfnamefont {T.~P.}\ \bibnamefont {Sotiriou}}\ and\ \bibinfo {author} {\bibfnamefont {V.}~\bibnamefont {Faraoni}},\ }\bibfield  {title} {\bibinfo {title} {{f(R) theories of gravity}},\ }\href {https://doi.org/10.1103/RevModPhys.82.451} {\bibfield  {journal} {\bibinfo  {journal} {Rev. Mod. Phys.}\ }\textbf {\bibinfo {volume} {82}},\ \bibinfo {pages} {451} (\bibinfo {year} {2010})},\ \Eprint {https://arxiv.org/abs/0805.1726} {arXiv:0805.1726 [gr-qc]} \BibitemShut {NoStop}%
\bibitem [{\citenamefont {Nojiri}\ and\ \citenamefont {Odintsov}(2011)}]{Nojiri:2010wj}%
  \BibitemOpen
  \bibfield  {author} {\bibinfo {author} {\bibfnamefont {S.}~\bibnamefont {Nojiri}}\ and\ \bibinfo {author} {\bibfnamefont {S.~D.}\ \bibnamefont {Odintsov}},\ }\bibfield  {title} {\bibinfo {title} {{Unified cosmic history in modified gravity: from F(R) theory to Lorentz non-invariant models}},\ }\href {https://doi.org/10.1016/j.physrep.2011.04.001} {\bibfield  {journal} {\bibinfo  {journal} {Phys. Rept.}\ }\textbf {\bibinfo {volume} {505}},\ \bibinfo {pages} {59} (\bibinfo {year} {2011})},\ \Eprint {https://arxiv.org/abs/1011.0544} {arXiv:1011.0544 [gr-qc]} \BibitemShut {NoStop}%
\bibitem [{\citenamefont {Nojiri}\ \emph {et~al.}(2005)\citenamefont {Nojiri}, \citenamefont {Odintsov},\ and\ \citenamefont {Sasaki}}]{Nojiri:2005vv}%
  \BibitemOpen
  \bibfield  {author} {\bibinfo {author} {\bibfnamefont {S.}~\bibnamefont {Nojiri}}, \bibinfo {author} {\bibfnamefont {S.~D.}\ \bibnamefont {Odintsov}},\ and\ \bibinfo {author} {\bibfnamefont {M.}~\bibnamefont {Sasaki}},\ }\bibfield  {title} {\bibinfo {title} {{Gauss-Bonnet dark energy}},\ }\href {https://doi.org/10.1103/PhysRevD.71.123509} {\bibfield  {journal} {\bibinfo  {journal} {Phys. Rev. D}\ }\textbf {\bibinfo {volume} {71}},\ \bibinfo {pages} {123509} (\bibinfo {year} {2005})},\ \Eprint {https://arxiv.org/abs/hep-th/0504052} {arXiv:hep-th/0504052} \BibitemShut {NoStop}%
\bibitem [{\citenamefont {Yang}\ \emph {et~al.}(2024)\citenamefont {Yang}, \citenamefont {Ren}, \citenamefont {Wang}, \citenamefont {Lu}, \citenamefont {Zhang}, \citenamefont {Cai},\ and\ \citenamefont {Saridakis}}]{Yang:2024kdo}%
  \BibitemOpen
  \bibfield  {author} {\bibinfo {author} {\bibfnamefont {Y.}~\bibnamefont {Yang}}, \bibinfo {author} {\bibfnamefont {X.}~\bibnamefont {Ren}}, \bibinfo {author} {\bibfnamefont {Q.}~\bibnamefont {Wang}}, \bibinfo {author} {\bibfnamefont {Z.}~\bibnamefont {Lu}}, \bibinfo {author} {\bibfnamefont {D.}~\bibnamefont {Zhang}}, \bibinfo {author} {\bibfnamefont {Y.-F.}\ \bibnamefont {Cai}},\ and\ \bibinfo {author} {\bibfnamefont {E.~N.}\ \bibnamefont {Saridakis}},\ }\bibfield  {title} {\bibinfo {title} {{Quintom cosmology and modified gravity after DESI 2024}},\ }\href@noop {} {\  (\bibinfo {year} {2024})},\ \Eprint {https://arxiv.org/abs/2404.19437} {arXiv:2404.19437 [astro-ph.CO]} \BibitemShut {NoStop}%
\bibitem [{\citenamefont {Cotton}(1899)}]{cotton1899varietes}%
  \BibitemOpen
  \bibfield  {author} {\bibinfo {author} {\bibfnamefont {{\'E}.}~\bibnamefont {Cotton}},\ }\bibfield  {title} {\bibinfo {title} {Sur les vari{\'e}t{\'e}s {\`a} trois dimensions},\ }\href@noop {} {\bibfield  {journal} {\bibinfo  {journal} {Annales de la Facult{\'e} des sciences de l'Universit{\'e} de Toulouse pour les sciences math{\'e}matiques et les sciences physiques}\ }\textbf {\bibinfo {volume} {1}},\ \bibinfo {pages} {385} (\bibinfo {year} {1899})}\BibitemShut {NoStop}%
\bibitem [{\citenamefont {Harada}(2021)}]{Harada:2021bte}%
  \BibitemOpen
  \bibfield  {author} {\bibinfo {author} {\bibfnamefont {J.}~\bibnamefont {Harada}},\ }\bibfield  {title} {\bibinfo {title} {{Emergence of the Cotton tensor for describing gravity}},\ }\href {https://doi.org/10.1103/PhysRevD.103.L121502} {\bibfield  {journal} {\bibinfo  {journal} {Phys. Rev. D}\ }\textbf {\bibinfo {volume} {103}},\ \bibinfo {pages} {L121502} (\bibinfo {year} {2021})},\ \Eprint {https://arxiv.org/abs/2105.09304} {arXiv:2105.09304 [gr-qc]} \BibitemShut {NoStop}%
\bibitem [{\citenamefont {Mantica}\ and\ \citenamefont {Molinari}(2023)}]{Mantica:2022flg}%
  \BibitemOpen
  \bibfield  {author} {\bibinfo {author} {\bibfnamefont {C.~A.}\ \bibnamefont {Mantica}}\ and\ \bibinfo {author} {\bibfnamefont {L.~G.}\ \bibnamefont {Molinari}},\ }\bibfield  {title} {\bibinfo {title} {{Codazzi tensors and their space-times and Cotton Gravity}},\ }\href {https://doi.org/10.1007/s10714-023-03106-7} {\bibfield  {journal} {\bibinfo  {journal} {Gen. Rel. Grav.}\ }\textbf {\bibinfo {volume} {55}},\ \bibinfo {pages} {62} (\bibinfo {year} {2023})},\ \Eprint {https://arxiv.org/abs/2210.06173} {arXiv:2210.06173 [gr-qc]} \BibitemShut {NoStop}%
\bibitem [{\citenamefont {Gogberashvili}\ and\ \citenamefont {Girgvliani}(2024)}]{Gogberashvili:2023wed}%
  \BibitemOpen
  \bibfield  {author} {\bibinfo {author} {\bibfnamefont {M.}~\bibnamefont {Gogberashvili}}\ and\ \bibinfo {author} {\bibfnamefont {A.}~\bibnamefont {Girgvliani}},\ }\bibfield  {title} {\bibinfo {title} {{General spherically symmetric solution of Cotton Gravity}},\ }\href {https://doi.org/10.1088/1361-6382/ad1781} {\bibfield  {journal} {\bibinfo  {journal} {Class. Quant. Grav.}\ }\textbf {\bibinfo {volume} {41}},\ \bibinfo {pages} {025010} (\bibinfo {year} {2024})},\ \Eprint {https://arxiv.org/abs/2308.03342} {arXiv:2308.03342 [gr-qc]} \BibitemShut {NoStop}%
\bibitem [{\citenamefont {Harada}(2022)}]{Harada:2022edl}%
  \BibitemOpen
  \bibfield  {author} {\bibinfo {author} {\bibfnamefont {J.}~\bibnamefont {Harada}},\ }\bibfield  {title} {\bibinfo {title} {{Cotton Gravity and 84 galaxy rotation curves}},\ }\href {https://doi.org/10.1103/PhysRevD.106.064044} {\bibfield  {journal} {\bibinfo  {journal} {Phys. Rev. D}\ }\textbf {\bibinfo {volume} {106}},\ \bibinfo {pages} {064044} (\bibinfo {year} {2022})},\ \Eprint {https://arxiv.org/abs/2209.04055} {arXiv:2209.04055 [gr-qc]} \BibitemShut {NoStop}%
\bibitem [{\citenamefont {Sussman}\ and\ \citenamefont {Najera}(2023{\natexlab{a}})}]{Sussman:2023wiw}%
  \BibitemOpen
  \bibfield  {author} {\bibinfo {author} {\bibfnamefont {R.~A.}\ \bibnamefont {Sussman}}\ and\ \bibinfo {author} {\bibfnamefont {S.}~\bibnamefont {Najera}},\ }\bibfield  {title} {\bibinfo {title} {{Cotton Gravity: the cosmological constant as spatial curvature}},\ }\href@noop {} {\  (\bibinfo {year} {2023}{\natexlab{a}})},\ \Eprint {https://arxiv.org/abs/2311.06744} {arXiv:2311.06744 [gr-qc]} \BibitemShut {NoStop}%
\bibitem [{\citenamefont {Sussman}\ and\ \citenamefont {Najera}(2023{\natexlab{b}})}]{Sussman:2023eep}%
  \BibitemOpen
  \bibfield  {author} {\bibinfo {author} {\bibfnamefont {R.~A.}\ \bibnamefont {Sussman}}\ and\ \bibinfo {author} {\bibfnamefont {S.}~\bibnamefont {Najera}},\ }\bibfield  {title} {\bibinfo {title} {{Exact solutions of Cotton Gravity in its Codazzi formulation}},\ }\href@noop {} {\  (\bibinfo {year} {2023}{\natexlab{b}})},\ \Eprint {https://arxiv.org/abs/2312.02115} {arXiv:2312.02115 [gr-qc]} \BibitemShut {NoStop}%
\bibitem [{\citenamefont {Mantica}\ and\ \citenamefont {Molinari}(2024)}]{Mantica:2023ssd}%
  \BibitemOpen
  \bibfield  {author} {\bibinfo {author} {\bibfnamefont {C.~A.}\ \bibnamefont {Mantica}}\ and\ \bibinfo {author} {\bibfnamefont {L.~G.}\ \bibnamefont {Molinari}},\ }\bibfield  {title} {\bibinfo {title} {{Friedmann equations in the Codazzi parametrization of Cotton and extended theories of gravity and the dark sector}},\ }\href {https://doi.org/10.1103/PhysRevD.109.044059} {\bibfield  {journal} {\bibinfo  {journal} {Phys. Rev. D}\ }\textbf {\bibinfo {volume} {109}},\ \bibinfo {pages} {044059} (\bibinfo {year} {2024})},\ \Eprint {https://arxiv.org/abs/2312.02784} {arXiv:2312.02784 [gr-qc]} \BibitemShut {NoStop}%
\bibitem [{\citenamefont {Cl\'ement}\ and\ \citenamefont {Nouicer}(2023)}]{Clement:2023tyx}%
  \BibitemOpen
  \bibfield  {author} {\bibinfo {author} {\bibfnamefont {G.}~\bibnamefont {Cl\'ement}}\ and\ \bibinfo {author} {\bibfnamefont {K.}~\bibnamefont {Nouicer}},\ }\bibfield  {title} {\bibinfo {title} {{Cotton Gravity is not predictive}},\ }\href@noop {} {\  (\bibinfo {year} {2023})},\ \Eprint {https://arxiv.org/abs/2312.17662} {arXiv:2312.17662 [gr-qc]} \BibitemShut {NoStop}%
\bibitem [{\citenamefont {Sussman}\ \emph {et~al.}(2024{\natexlab{a}})\citenamefont {Sussman}, \citenamefont {Mantica}, \citenamefont {Molinari},\ and\ \citenamefont {N\'ajera}}]{Sussman:2024iwk}%
  \BibitemOpen
  \bibfield  {author} {\bibinfo {author} {\bibfnamefont {R.~A.}\ \bibnamefont {Sussman}}, \bibinfo {author} {\bibfnamefont {C.~A.}\ \bibnamefont {Mantica}}, \bibinfo {author} {\bibfnamefont {L.~G.}\ \bibnamefont {Molinari}},\ and\ \bibinfo {author} {\bibfnamefont {S.}~\bibnamefont {N\'ajera}},\ }\bibfield  {title} {\bibinfo {title} {{Response to a critique of ``Cotton Gravity''}},\ }\href@noop {} {\  (\bibinfo {year} {2024}{\natexlab{a}})},\ \Eprint {https://arxiv.org/abs/2401.10479} {arXiv:2401.10479 [gr-qc]} \BibitemShut {NoStop}%
\bibitem [{\citenamefont {Cl\'ement}\ and\ \citenamefont {Nouicer}(2024)}]{Clement:2024pjl}%
  \BibitemOpen
  \bibfield  {author} {\bibinfo {author} {\bibfnamefont {G.}~\bibnamefont {Cl\'ement}}\ and\ \bibinfo {author} {\bibfnamefont {K.}~\bibnamefont {Nouicer}},\ }\bibfield  {title} {\bibinfo {title} {{Farewell to Cotton Gravity}},\ }\href@noop {} {\  (\bibinfo {year} {2024})},\ \Eprint {https://arxiv.org/abs/2401.16008} {arXiv:2401.16008 [gr-qc]} \BibitemShut {NoStop}%
\bibitem [{\citenamefont {Sussman}\ \emph {et~al.}(2024{\natexlab{b}})\citenamefont {Sussman}, \citenamefont {Mantica}, \citenamefont {Molinari},\ and\ \citenamefont {N\'ajera}}]{Sussman:2024qsg}%
  \BibitemOpen
  \bibfield  {author} {\bibinfo {author} {\bibfnamefont {R.~A.}\ \bibnamefont {Sussman}}, \bibinfo {author} {\bibfnamefont {C.~A.}\ \bibnamefont {Mantica}}, \bibinfo {author} {\bibfnamefont {L.~G.}\ \bibnamefont {Molinari}},\ and\ \bibinfo {author} {\bibfnamefont {S.}~\bibnamefont {N\'ajera}},\ }\bibfield  {title} {\bibinfo {title} {{Second response to the critique of ``Cotton Gravity''}},\ }\href@noop {} {\  (\bibinfo {year} {2024}{\natexlab{b}})},\ \Eprint {https://arxiv.org/abs/2402.01992} {arXiv:2402.01992 [gr-qc]} \BibitemShut {NoStop}%
\bibitem [{\citenamefont {Zhang}\ \emph {et~al.}(2023)\citenamefont {Zhang}, \citenamefont {Li}, \citenamefont {Li}, \citenamefont {Yang}, \citenamefont {Zhang}, \citenamefont {Cai}, \citenamefont {Fang},\ and\ \citenamefont {Feng}}]{Zhang:2021ecp}%
  \BibitemOpen
  \bibfield  {author} {\bibinfo {author} {\bibfnamefont {D.}~\bibnamefont {Zhang}}, \bibinfo {author} {\bibfnamefont {J.-R.}\ \bibnamefont {Li}}, \bibinfo {author} {\bibfnamefont {J.}~\bibnamefont {Li}}, \bibinfo {author} {\bibfnamefont {J.}~\bibnamefont {Yang}}, \bibinfo {author} {\bibfnamefont {Y.}~\bibnamefont {Zhang}}, \bibinfo {author} {\bibfnamefont {Y.-F.}\ \bibnamefont {Cai}}, \bibinfo {author} {\bibfnamefont {W.}~\bibnamefont {Fang}},\ and\ \bibinfo {author} {\bibfnamefont {C.}~\bibnamefont {Feng}},\ }\bibfield  {title} {\bibinfo {title} {{Future prospects on constraining neutrino cosmology with the Ali CMB Polarization Telescope}},\ }\href {https://doi.org/10.3847/1538-4357/acbe45} {\bibfield  {journal} {\bibinfo  {journal} {Astrophys. J.}\ }\textbf {\bibinfo {volume} {946}},\ \bibinfo {pages} {32} (\bibinfo {year} {2023})},\ \Eprint {https://arxiv.org/abs/2112.10539} {arXiv:2112.10539 [astro-ph.CO]} \BibitemShut {NoStop}%
\bibitem [{\citenamefont {Brust}\ \emph {et~al.}(2013)\citenamefont {Brust}, \citenamefont {Kaplan},\ and\ \citenamefont {Walters}}]{Brust:2013ova}%
  \BibitemOpen
  \bibfield  {author} {\bibinfo {author} {\bibfnamefont {C.}~\bibnamefont {Brust}}, \bibinfo {author} {\bibfnamefont {D.~E.}\ \bibnamefont {Kaplan}},\ and\ \bibinfo {author} {\bibfnamefont {M.~T.}\ \bibnamefont {Walters}},\ }\bibfield  {title} {\bibinfo {title} {{New light species and the CMB}},\ }\href {https://doi.org/10.1007/JHEP12(2013)058} {\bibfield  {journal} {\bibinfo  {journal} {JHEP}\ }\textbf {\bibinfo {volume} {12}},\ \bibinfo {pages} {058}},\ \Eprint {https://arxiv.org/abs/1303.5379} {arXiv:1303.5379 [hep-ph]} \BibitemShut {NoStop}%
\bibitem [{\citenamefont {Marsh}(2016)}]{Marsh:2015xka}%
  \BibitemOpen
  \bibfield  {author} {\bibinfo {author} {\bibfnamefont {D.~J.~E.}\ \bibnamefont {Marsh}},\ }\bibfield  {title} {\bibinfo {title} {{Axion cosmology}},\ }\href {https://doi.org/10.1016/j.physrep.2016.06.005} {\bibfield  {journal} {\bibinfo  {journal} {Phys. Rept.}\ }\textbf {\bibinfo {volume} {643}},\ \bibinfo {pages} {1} (\bibinfo {year} {2016})},\ \Eprint {https://arxiv.org/abs/1510.07633} {arXiv:1510.07633 [astro-ph.CO]} \BibitemShut {NoStop}%
\bibitem [{\citenamefont {Komatsu}(2022)}]{Komatsu:2022nvu}%
  \BibitemOpen
  \bibfield  {author} {\bibinfo {author} {\bibfnamefont {E.}~\bibnamefont {Komatsu}},\ }\bibfield  {title} {\bibinfo {title} {{New physics from the polarized light of the cosmic microwave background}},\ }\href {https://doi.org/10.1038/s42254-022-00452-4} {\bibfield  {journal} {\bibinfo  {journal} {Nature Rev. Phys.}\ }\textbf {\bibinfo {volume} {4}},\ \bibinfo {pages} {452} (\bibinfo {year} {2022})},\ \Eprint {https://arxiv.org/abs/2202.13919} {arXiv:2202.13919 [astro-ph.CO]} \BibitemShut {NoStop}%
\bibitem [{\citenamefont {Abazajian}\ \emph {et~al.}(2009)\citenamefont {Abazajian} \emph {et~al.}}]{SDSS:2008tqn}%
  \BibitemOpen
  \bibfield  {author} {\bibinfo {author} {\bibfnamefont {K.~N.}\ \bibnamefont {Abazajian}} \emph {et~al.} (\bibinfo {collaboration} {SDSS}),\ }\bibfield  {title} {\bibinfo {title} {{The Seventh Data Release of the Sloan Digital Sky Survey}},\ }\href {https://doi.org/10.1088/0067-0049/182/2/543} {\bibfield  {journal} {\bibinfo  {journal} {Astrophys. J. Suppl.}\ }\textbf {\bibinfo {volume} {182}},\ \bibinfo {pages} {543} (\bibinfo {year} {2009})},\ \Eprint {https://arxiv.org/abs/0812.0649} {arXiv:0812.0649 [astro-ph]} \BibitemShut {NoStop}%
\bibitem [{\citenamefont {Sakr}\ and\ \citenamefont {Martinelli}(2022)}]{Sakr:2021ylx}%
  \BibitemOpen
  \bibfield  {author} {\bibinfo {author} {\bibfnamefont {Z.}~\bibnamefont {Sakr}}\ and\ \bibinfo {author} {\bibfnamefont {M.}~\bibnamefont {Martinelli}},\ }\bibfield  {title} {\bibinfo {title} {{Cosmological constraints on sub-horizon scales modified gravity theories with MGCLASS II}},\ }\href {https://doi.org/10.1088/1475-7516/2022/05/030} {\bibfield  {journal} {\bibinfo  {journal} {JCAP}\ }\textbf {\bibinfo {volume} {05}},\ \bibinfo {pages} {030}},\ \Eprint {https://arxiv.org/abs/2112.14175} {arXiv:2112.14175 [astro-ph.CO]} \BibitemShut {NoStop}%
\bibitem [{\citenamefont {Blas}\ \emph {et~al.}(2011)\citenamefont {Blas}, \citenamefont {Lesgourgues},\ and\ \citenamefont {Tram}}]{Blas:2011rf}%
  \BibitemOpen
  \bibfield  {author} {\bibinfo {author} {\bibfnamefont {D.}~\bibnamefont {Blas}}, \bibinfo {author} {\bibfnamefont {J.}~\bibnamefont {Lesgourgues}},\ and\ \bibinfo {author} {\bibfnamefont {T.}~\bibnamefont {Tram}},\ }\bibfield  {title} {\bibinfo {title} {{The Cosmic Linear Anisotropy Solving System (CLASS) II: approximation schemes}},\ }\href {https://doi.org/10.1088/1475-7516/2011/07/034} {\bibfield  {journal} {\bibinfo  {journal} {JCAP}\ }\textbf {\bibinfo {volume} {07}},\ \bibinfo {pages} {034}},\ \Eprint {https://arxiv.org/abs/1104.2933} {arXiv:1104.2933 [astro-ph.CO]} \BibitemShut {NoStop}%
\bibitem [{\citenamefont {Brinckmann}\ and\ \citenamefont {Lesgourgues}(2019)}]{Brinckmann:2018cvx}%
  \BibitemOpen
  \bibfield  {author} {\bibinfo {author} {\bibfnamefont {T.}~\bibnamefont {Brinckmann}}\ and\ \bibinfo {author} {\bibfnamefont {J.}~\bibnamefont {Lesgourgues}},\ }\bibfield  {title} {\bibinfo {title} {{MontePython 3: boosted MCMC sampler and other features}},\ }\href {https://doi.org/10.1016/j.dark.2018.100260} {\bibfield  {journal} {\bibinfo  {journal} {Phys. Dark Univ.}\ }\textbf {\bibinfo {volume} {24}},\ \bibinfo {pages} {100260} (\bibinfo {year} {2019})},\ \Eprint {https://arxiv.org/abs/1804.07261} {arXiv:1804.07261 [astro-ph.CO]} \BibitemShut {NoStop}%
\bibitem [{\citenamefont {Audren}\ \emph {et~al.}(2013)\citenamefont {Audren}, \citenamefont {Lesgourgues}, \citenamefont {Benabed},\ and\ \citenamefont {Prunet}}]{Audren:2012wb}%
  \BibitemOpen
  \bibfield  {author} {\bibinfo {author} {\bibfnamefont {B.}~\bibnamefont {Audren}}, \bibinfo {author} {\bibfnamefont {J.}~\bibnamefont {Lesgourgues}}, \bibinfo {author} {\bibfnamefont {K.}~\bibnamefont {Benabed}},\ and\ \bibinfo {author} {\bibfnamefont {S.}~\bibnamefont {Prunet}},\ }\bibfield  {title} {\bibinfo {title} {{Conservative constraints on early cosmology: an illustration of the Monte Python cosmological parameter inference code}},\ }\href {https://doi.org/10.1088/1475-7516/2013/02/001} {\bibfield  {journal} {\bibinfo  {journal} {JCAP}\ }\textbf {\bibinfo {volume} {02}},\ \bibinfo {pages} {001}},\ \Eprint {https://arxiv.org/abs/1210.7183} {arXiv:1210.7183 [astro-ph.CO]} \BibitemShut {NoStop}%
\bibitem [{\citenamefont {Aghanim}\ \emph {et~al.}(2020{\natexlab{b}})\citenamefont {Aghanim} \emph {et~al.}}]{Planck:2019nip}%
  \BibitemOpen
  \bibfield  {author} {\bibinfo {author} {\bibfnamefont {N.}~\bibnamefont {Aghanim}} \emph {et~al.} (\bibinfo {collaboration} {Planck}),\ }\bibfield  {title} {\bibinfo {title} {{Planck 2018 results. V. CMB power spectra and likelihoods}},\ }\href {https://doi.org/10.1051/0004-6361/201936386} {\bibfield  {journal} {\bibinfo  {journal} {Astron. Astrophys.}\ }\textbf {\bibinfo {volume} {641}},\ \bibinfo {pages} {A5} (\bibinfo {year} {2020}{\natexlab{b}})},\ \Eprint {https://arxiv.org/abs/1907.12875} {arXiv:1907.12875 [astro-ph.CO]} \BibitemShut {NoStop}%
\bibitem [{Note1()}]{Note1}%
  \BibitemOpen
  \bibinfo {note} {\protect \url {https://pla.esac.esa.int/\#cosmology}}\BibitemShut {NoStop}%
\bibitem [{\citenamefont {Garcia-Arroyo}\ \emph {et~al.}(2020)\citenamefont {Garcia-Arroyo}, \citenamefont {Cervantes-Cota}, \citenamefont {Nucamendi},\ and\ \citenamefont {Aviles}}]{Garcia-Arroyo:2020iou}%
  \BibitemOpen
  \bibfield  {author} {\bibinfo {author} {\bibfnamefont {G.}~\bibnamefont {Garcia-Arroyo}}, \bibinfo {author} {\bibfnamefont {J.~L.}\ \bibnamefont {Cervantes-Cota}}, \bibinfo {author} {\bibfnamefont {U.}~\bibnamefont {Nucamendi}},\ and\ \bibinfo {author} {\bibfnamefont {A.}~\bibnamefont {Aviles}},\ }\bibfield  {title} {\bibinfo {title} {{Effects of dark energy anisotropic stress on the matter power spectrum}},\ }\href {https://doi.org/10.1016/j.dark.2020.100668} {\bibfield  {journal} {\bibinfo  {journal} {Phys. Dark Univ.}\ }\textbf {\bibinfo {volume} {30}},\ \bibinfo {pages} {100668} (\bibinfo {year} {2020})},\ \Eprint {https://arxiv.org/abs/2004.13917} {arXiv:2004.13917 [astro-ph.CO]} \BibitemShut {NoStop}%
\end{thebibliography}%

\end{document}